\documentclass[final]{hubsarxiv}

\usepackage{tabularx}
\usepackage{multirow}
\usepackage{longtable}
\usepackage{rotating}
\newenvironment{surveyblock}{\par\medskip\noindent\rule{\textwidth}{0.4pt}\par\small\setlength{\parskip}{3pt}\setlength{\parindent}{0pt}}{\par\noindent\rule{\textwidth}{0.4pt}\par\medskip}

\title{Human-AI Collaboration in Large Language Model-Integrated Building Energy Management Systems: The Role of User Domain Knowledge and AI Literacy}
\shorttitle{Human-AI collaboration in LLM-integrated BEMS}
\shortauthors{Jung, Jeon and Babon-Ayeng}
\author{Wooyoung Jung\affmark{a,}\corrmark\orcid{0000-0003-2746-4983} \and Kahyun Jeon\affmark{b} \and Prosper Babon-Ayeng\affmark{a}}
\affil{\affmark{a}Department of Civil and Architectural Engineering and Mechanics, University of Arizona, Tucson, AZ, USA\\
\affmark{b}Illinois Institute of Technology, Chicago, IL, USA}
\corresponding{wooyoung@arizona.edu}
\paperno{2026-002}
\paperdate{February 2026}
\versionnote{v1 (2026-02-18): first release. v2 (2026-11-05): re-set in the HUBS Lab paper layout; content unchanged.}
\license{CC BY-NC-ND 4.0}
\abstract{This study aimed to comprehend how user domain knowledge and artificial intelligence (AI) literacy impact the effective use of human-AI interactive building energy management system (BEMS). While prior studies have investigated the potential of integrating large language models (LLMs) into BEMS or building energy modeling, very few studies have examined how user interact with such systems. We conducted a systematic role-playing experiment, where 85 human subjects interacted with an advanced generative pre-trained transformer (OpenAI GPT-4o). Participants were tasked with identifying the top five behavioral changes that could reduce home energy use with the GPT model that functioned as an LLM-integrated BEMS. Then, the collected prompt-response data and participant conclusions were analyzed using an analytical framework that hierarchically assessed and scored human-AI interactions and their home energy analysis approaches. Also, participants were classified into four groups based on their self-evaluated domain knowledge of building energy use and AI literacy, and Kruskal-Wallis H tests with post-hoc pairwise comparisons were conducted across 20 quantifiable metrics. Key takeaways include: most participants employed concise prompts (median: 16.2 words) and relied heavily on GPT's analytical capabilities; and notably, only 1 of 20 metrics -- appliance identification rate -- showed statistically significant group differences (p=0.037), driven by AI literacy rather than domain knowledge, suggesting an equalizing effect of LLMs across expertise levels. This study provides foundational insights into human-AI collaboration dynamics and promising development directions in the context of LLM-integrated BEMS and contributes to realizing human-centric LLM-integrated energy systems.}

\begin{document}
\maketitle

\section{Introduction}

Building energy use patterns are uniquely and stochastically shaped by occupants' diverse interactions with appliances and building systems such as heating, ventilation, and air-conditioning (HVAC), lighting, and plumbing. Each occupant holds their own preferences and needs, which determine the operation, scheduling, and optimization of these systems -- often balancing comfort, budgetary constraints and energy saving goals \cite{r1}. Despite this heterogeneity, at present, occupants receive limited feedback on their energy use. Monthly energy bills serve primarily as `financial consequences' of their behavior, panelizing inefficient energy use without providing guidance or actionable insights for improvement. Even though the widespread installation of smart meters has enabled utility-provided mobile apps to provide users hourly or daily tracking of aggregated energy consumption, achieving a representing progress in information accessibility (e.g., \cite{r2,r3}), users have limited opportunities to interaction with building energy management systems (BEMSs). Users are often confined to predefined software features that seldom offer interactive capabilities for knowledge acquisition or data analysis. For instance, occupants often hesitate to enroll in time-of-use (TOU) rate programs -- to encourage peak load reduction and load shifting through behavioral changes (e.g. \cite{r4,r5,r6}) -- due to a lack of detailed, personalized information of how subscribing TOU rates may affect their specific energy bills.

\begin{figure}[tbp]
\centering
\includegraphics[width=0.69\textwidth]{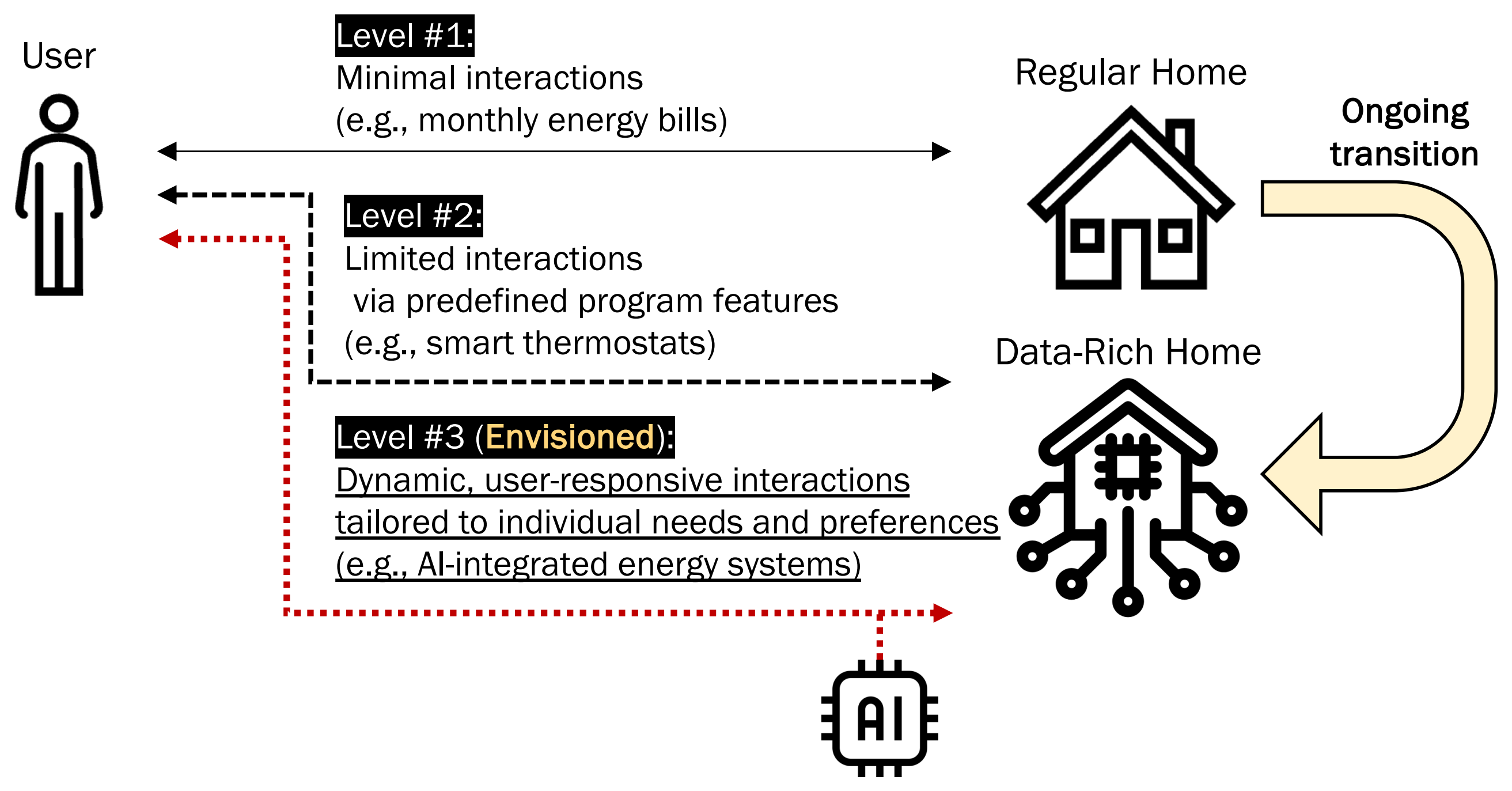}
\caption{Current and envisioned human building interactions in the context of BEMS}\label{fig:1}
\end{figure}

In recent years, a few studies have proposed a large language model (LLM)-integrated BEMSs, which can address this long-lasting limitation \cite{r7,r8,r9}. These studies showed the potential of human-centered interactions through their systems, where occupants can readily access and generate information from multiple home energy data sources, receive personalized recommendations for energy efficiency improvements or behavioral changes, and streamline control over appliances and building systems using natural language, as illustrated in Figure~\ref{fig:1} as Level \#3. As their primary focus was to evaluate their systems' feasibility, their experimental settings tested various user queries, covering a wide range of building energy management (BEM)-related tasks such as ``How much energy did I use last month?'' or ``Set the temperature setpoint as 24°C''. The major limitation of these single-turn query-response trials is that they do not reveal how users actually collaborate with LLMs through iterative, multi-turn dialogues to build understanding and arrive at actionable energy-saving decisions. In practice, effective use of LLM-integrated BEMS requires users to interpret responses, refine their inquiries, evaluate suggested strategies against personal constraints, and synthesize information across multiple exchanges.

We anticipate two primary challenges when these LLM-integrated BEMS are deployed in practice from the users' perspective: (1) \emph{the complexity of building energy management,} which involves multiple appliances and building systems (e.g., HVAC, lighting, plumbing) -- each with distinct control mechanisms and engineering principles -- operating within a system-of-systems structure. This complexity has further intensified in recent years, with utilities introducing new rate structures, such as TOU rates, to integrate renewable energy resources and promote demand response (DR) strategies; (2) \emph{the requirements for AI literacy}, which extend beyond merely crafting well-structured prompts \cite{r10}. AI literacy comprises a set of competencies that enables users to critically evaluate AI technologies, communicate and collaborate effectively with AI systems, and utilize AI as a tool to achieve specific goals \cite{r11}. These competencies include providing sufficient detail, offering contextual framing (e.g., assigning GPT an appropriate expert role), and engaging in iterative questioning and reflection to improve the relevance, clarity, and personalization of the interaction. Accordingly, without adequately addressing these challenges, households may struggle to harness the full potential of LLM-integrated BEMS to access, interpret, and act on critical building energy management (BEM) information or may encounter system responses that are too generic, misaligned with household needs, or insufficiently contextualized to support informed energy decisions.

Prior research in broader human-AI interaction contexts suggests that these two factors can significantly shape the quality of collaboration. Studies on AI-assisted decision-making have reported that users' domain knowledge influences how they interpret, evaluate, and calibrate their reliance on AI-generated recommendations \cite{r12}. In addition, research on AI literacy has indicated that users' ability to articulate effective prompts, engage in iterative dialogue, and critically evaluate AI outputs can affect the outcomes of human-AI collaboration \cite{r13}. However, whether and how these factors operate in the context of LLM-integrated BEMS, where users engage in multi-turn, goal-oriented conversations involving complex, domain-specific energy data, remains an open question.

Consequently, this study aimed to examine human-AI interactions in the context of LLM-integrated BEMS. Our research question was developed as follows:

\begin{itemize}
\item
  \emph{\textbf{Which characteristics}} -- \emph{\textbf{particularly domain knowledge and AI literacy}} -- \emph{\textbf{enable households to fully benefit from LLM-enhanced building energy management?}}
\end{itemize}

To accomplish the research goal and address the research question, we designed a role-playing experiment in which participants were tasked with identifying five feasible behavioral changes to reduce household energy bills through interactions with an advanced generative pretrained transformer (GPT) model, with no time constraints (see Section 3 for more details). Our hypotheses were as follows:

\begin{enumerate}
\def\labelenumi{\arabic{enumi}.}
\item
  Participants with limited domain knowledge may rely heavily on the LLM's responses, potentially resulting in brief interactions.
\item
  Participants with strong domain knowledge may engage in extended interactions with the tool to validate LLM's suggestions, ultimately arriving at reasonably accurate conclusions.
\item
  Participants paired with high AI literacy would compensate their limited domain knowledge by effectively interacting with GPT to extract relevant insights, formulate well-structured prompts, and critically assess AI-generated recommendations.
\item
  Excessively verbose GPT responses may negatively affect user agency and limit iterative dialogue, especially in participants with lower AI literacy.
\item
  Participants would perceive this LLM-based BEMS as useful and even educational, providing an accessible and engaging opportunity to enhance their understanding of home energy systems while simultaneously identifying feasible energy-saving strategies.
\end{enumerate}

Using a prescreening survey, we utilized a stratified sampling approach to recruit participants from different user groups based on their domain knowledge and AI literacy. A total of 85 participants conducted the role-playing experiment and then participated in a post-experiment survey. We then analyzed the interactions of 85 participants with an LLM tool -- specifically, ChatGPT-4o by OpenAI -- to investigate the relationships among domain knowledge, AI literacy, and interaction outcomes.

The key contributions of this study are threefold: (1) a systematic empirical investigation of how user domain knowledge and AI literacy shape human-AI interaction in LLM-integrated BEMS, using a 2×2 factorial design with 85 participants and non-parametric inferential statistical analyses; (2) the introduction of the SCALE framework and two complementary metric sets (interaction volume and conversational reasoning) for analyzing the structure, depth, and quality of human-AI dialogues; and (3) evidence-based design implications for developing more effective LLM-integrated energy management platforms, including scaffolded interaction designs, interpretive support features, and adaptive output mechanisms.

\section{Literature Review}

\subsection{Use of Language Models in Home/Building Energy Management}

\textbf{Pre-LLM.} Despite several decades of natural language processing (NLP) development, its application in BEM remains limited, particularly for text data processing and analysis. One common use case is building code compliance checking. Fuchs \cite{r14} conducted a comprehensive and systematic review of this domain, covering 42 research articles published between 2000 and 2020. NLP was utilized to process building code documents, classify texts, extract key features and information. These outputs were integrated with model data extracted from building information models to assess compliances. In a similar context, Khanuja \cite{r15} used NLP to process energy efficiency measures data -- often provided in non-standardized text formats -- to ensure consistency and standardization. Lai et al. \cite{r16} and Hong et al. \cite{r17} used fundamental NLP techniques such as tokenization, stop word removal, and lemmatization to extract relevant information, including energy conservation measures. Olanrewaju et al. \cite{r18} applied NLP-based document similarity techniques -- such as cosine similarity and term frequency-inverse document frequency (TF-IDF) \cite{r19} -- to evaluate technical manuals associated with green building certification systems, such as Leadership in Energy and Environmental Design).

The limited application in NLP have not significantly enhanced user interactions in BEMS, mainly enabling basic-level communication features. For example, Kim et al. \cite{r20} developed a task-oriented voice user interface (VUI), facilitating the interaction between users and their cloud-based eco-feedback and gaming platform called MySmartE. This VUI utilized Amazon's Alexa Skills Kit, which enables developers to create conversational interfaces for appliances. Users could control their thermostats and review their energy savings using the VUI. However, as noted by the authors, the VUI was less frequently used than the visual user interface and was mainly employed for simple control tasks or information queries. Kumar et al. \cite{r21} presented an approach that allowed users to monitor indoor environmental quality (e.g., air temperature, humidity) through voice-activated controls. However, their system's interaction remained unidirectional, limited to command execution, without supporting conversational exchanges through which users could inquire about their energy use patterns or receive contextual feedback.

\textbf{Post-LLM.} Since the public release of LLM tools in early 2020s, a growing number of studies have applied these technologies to the field of BEMS. To date, the primary focus of these studies has been on feasibility assessment -- evaluating whether LLMs possessed sufficient domain knowledge for BEMS (e.g., \cite{r22}) and demonstrating their potential in specific use cases such as automated energy waste detection \cite{r23}, urban energy modeling \cite{r24}, fault diagnosis \cite{r25,r26,r27}, and energy management optimization \cite{r28} -- or on technical integration strategies such as domain-specific fine-tuning and retrieval augmented generation (RAG) \cite{r29,r30}. These works have largely centered on \textbf{system capabilities}, including the types of questions LLMs can learn and answer, the accuracy of their energy-related outputs, and general recommendations for how such tools might be deployed in future smart home and building contexts.

This trend persisted in the studies that focused on user aspects of LLM integration into BEMS. Specifically, He and Jazizadeh \cite{r7}, \cite{r8} employed a pre-built platform, OpenAI Assistants Application Programming Interface, to build their agentic system to incorporate tools (code interpreter, knowledge retrieval, and function calling) and to assess the feasibility of enhancing building energy efficiency and management through LLM-integrated BEMS. They evaluated 95 queries involving information requests, analysis and suggestions, and connections to external devices utilizing various prompt types. Rey-Jouanchicot et al. \cite{r9} proposed an LLM-based smart home architecture that leveraged user preferences to proactively select device actions in response to home events, such as adjusting lighting when a user gets out of bed. Their system was evaluated on 11 predefined single-turn scenarios, where each home state triggered a single LLM-generated action.

In contrast, this study shifts the focus from system feasibility to the \textbf{human and user side of LLM integration into BEMS}. While prior work has established that LLMs possess sufficient analytical capabilities for building energy applications, which demonstrates system-level competence does not ensure that users will effectively leverage these capabilities in practice. The quality of human-AI collaboration ultimately depends on how users interact with, interpret, and act upon AI-generated outputs. Rather than evaluating what LLMs can do, this research examines how actual users with varying backgrounds interact with a GenAI tool to explore BEM. Through a systematic human subject experiment and a multi-layered analysis of participant-GPT interactions, this study aims to provide insight into the practical dynamics of human-AI collaboration in the context of BEMS.

\subsection{Domain Knowledge and AI Literacy in Human-AI Collaboration}

As noted in the Introduction, prior research in broader human-AI interaction contexts has identified domain knowledge and AI literacy as factors that can shape the quality of collaboration between users and AI systems. This section reviews the empirical evidence underlying these two factors in greater detail.

\textbf{Domain knowledge}. A growing body of research has examined how users' prior knowledge of a task domain affects their behavior when working with AI-assisted decision-support systems. Dikmen and Burns \cite{r12} investigated the role of domain knowledge in the context of AI-assisted peer-to-peer lending decisions and found that participants who possessed relevant domain knowledge exhibited more calibrated trust compared with those who lacked such knowledge. Nourani et al. \cite{r31} further demonstrated that domain experts were able to dynamically adjust their trust based on the AI system's observed performance, whereas novice users exhibited persistent over-reliance regardless of the system's accuracy. In a study on human-AI team performance, Inkpen et al. \cite{r32} reported that highly proficient users were generally able to discern when to follow AI recommendations and when to override them, contributing to improved complementary team performance. Conversely, Bansal et al. \cite{r33} revealed that without sufficient domain knowledge to independently assess the correctness of AI outputs, users tended to follow AI-generated explanations indiscriminately. Collectively, these findings suggest that domain knowledge can serve as a moderating factor in how users process, evaluate, and act upon AI-generated information, though the specific manifestation of this effect may vary across task contexts.

\textbf{AI literacy.} In parallel, a separate line of research has examined how users' familiarity with AI tools, particularly their ability to construct effective prompts, engage in iterative refinement, and critically assess AI outputs \cite{r34,r35}, influences the quality of human-AI collaboration. Knoth et al. \cite{r13} investigated prompt engineering behaviors among users with different levels of AI literacy and found that users with lower AI literacy tended to construct prompts that mimicked human-to-human instructions, whereas those with higher AI literacy employed more structured, iterative, and strategically refined prompting approaches. Kim et al. \cite{r36} observed a similar pattern in an educational setting: students with high AI literacy engaged in collaborative, context-rich interactions with GenAI and produced significantly better writing outcomes than those with low AI literacy, who relied on generic, one-directional prompts with minimal iteration. Liu et al. \cite{r37} further demonstrated that AI literacy was a significant predictor of job performance in workplace settings.

\textbf{Synthesis}. While these studies provide converging evidence that both domain knowledge and AI literacy can shape human-AI interaction dynamics, a few important caveats should be noted. First, much of the evidence on domain knowledge originates from classification or decision-support tasks in which users make discrete judgments, such as accepting or rejecting a single AI recommendation. Similarly, findings on AI literacy have been drawn primarily from educational and workplace contexts. Whether these patterns hold in the context of LLM-integrated BEMS, where users engage in multi-turn, goal-oriented conversations to interpret complex, domain-specific energy data and arrive at actionable decisions, has not been empirically tested. Second, the interaction between these two factors remains underexplored; it is plausible that domain knowledge and AI literacy may compensate for, reinforce, or even diminish each other's effects depending on the task context. These gaps motivate the present study's research design, which systematically examines their individual and combined effects on user-LLM interaction in the context of BEM.

\section{Methodology}

\subsection{Overview}

Grounded in the evidence reviewed in Section 2, we designed a controlled experiment to examine how these two factors influence user-LLM collaboration in the context of BEM. Specifically, we adopted a role-playing experimental design -- an established methodology for studying human decision-making \cite{r38} -- in which participants assumed the role of a homeowner and were tasked with identifying the top five feasible behavioral changes to reduce households energy bills through interactions with OpenAI's ChatGPT-4o model. This approach enabled systematic manipulation of the two independent variables while preserving a realistic task context grounded in actual appliance-level energy data.

\begin{figure}[tbp]
\centering
\includegraphics[width=1.00\textwidth]{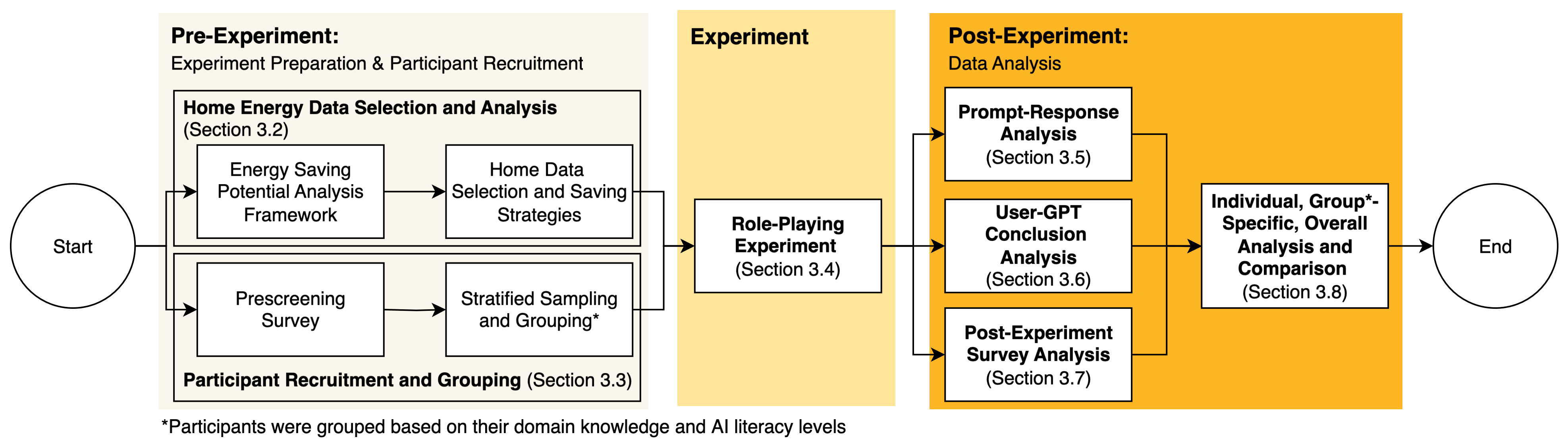}
\caption{Methodology: An overview}\label{fig:2}
\end{figure}

Figure~\ref{fig:2} presents a flowchart outlining the preparation, execution, and analysis phases of the study, with each component corresponding to the subsections that follow. This study was approved by the University of Arizona's Institutional Review Board (IRB) in January, 2025 (STUDY00005726) as a minimal risk experiment, and data collection was conducted from February to May 2025.

\subsection{Pre-Experiment: Home Energy Use Data Selection and Analysis}

This part of the research aimed to establish a systematic analysis framework to identify home energy saving potential using appliance-level energy use data (Section 3.2.1) and create reference conclusions for use in the role-playing experiment (Section 3.2.2). To set up the experimental environment, we utilized a single home case as the foundation. This approach was chosen to maintain consistency and focus our analysis on participants' interactions with the GPT model.

\subsubsection{Energy Saving Potential Analysis Framework}

As our energy use analysis framework to draw reference conclusions, we developed a set of energy-saving potential metrics by revising and extending the user DR potential scoring approach shown in \cite{r39}. Specifically, our framework introduced two key improvements: (1) it differentiated between on-peak and off-peak timeframes and (2) it introduced two new metrics, appliance use flexibility and comfort association with appliance, to more comprehensively characterize appliance-level energy-saving potential, as detailed below:

\begin{itemize}
\item
  \textbf{Appliance Energy Use:} We considered the mean and variance of energy consumption for an appliance or circuit when activated. To assess energy-saving potential, a higher mean energy use combined with lower variance suggests greater potential for energy savings.
\item
  \textbf{Appliance Use Frequency:} This metric quantifies how often a household uses a specific appliance or circuit. It provides insight into whether usage can be reduced within two distinct timeframes (on- and off-peak).
\item
  \textbf{Appliance Use Flexibility:} This metric assesses how rigidly or flexibly an appliance or circuit is used. For example, when a user consistently operates an appliance -- such as an oven -- during a fixed time window (e.g., 11:30 am to 12:00 pm), it suggests a habitual routine that may be difficult to modify. This behavioral rigidity, particularly paired with high use frequency, indicates a low potential for energy savings.
\item
  \textbf{Comfort Association with Appliance:} This Boolean metric assesses whether a specific appliance influences user comfort (True or False). A behavioral change is considered feasible only if it reduces energy consumption without significantly impacting comfort or health. For example, altering a thermostat's temperature setpoint by 5°F (2.8°C) might not be practical, as such changes could alter occupants' thermal sensations and lead to thermal discomfort.
\end{itemize}

The following equations are the details of our proposed metrics above. Table~\ref{tab:1} shares the basic notations and their descriptions that we utilized to establish all metrics.

\begin{table}[tbp]
\centering
\caption{Basic notations and their descriptions}\label{tab:1}
\begin{tabular}{@{}ll@{}}
\toprule
Notation & Description \\
\midrule
$i$ & Appliance or circuit type \\
$t$ & Time within the day or timeframe (e.g., hour) \\
$D$ & The total number of days in the dataset \\
$P_{i,t}$ & Power consumption of appliance or circuit $i$ at time $t$ on day $j$ \\
$T_{i}$ & Power threshold specific to the appliance or circuit $i$ \\
\bottomrule
\end{tabular}
\end{table}

A binary indicator function \(U_{i,t}\) (Equation (1)) to detect whether the appliance or circuit \(i\) was in use at time \(t\):

\begin{equation}\tag{1}
U_{i,t} = \left\{ \begin{array}{r}
1 \\
0
\end{array} \right.\ \begin{matrix}
\text{if } P_{i,t} \geq T_{i} \\
\text{if } P_{i,t} < T_{i}
\end{matrix}
\end{equation}

In this study, we applied adaptive thresholds after examining the power consumptions of active and standby modes in each appliance and circuit. We referenced a couple of articles \cite{r40,r41} as our starting points and heuristically refined our thresholds.

The hourly use frequency (\(F_{i,H_{j}}\)) was the sum of binary indicator values (\(U_{i,t}\)) for all 15-minute intervals within an hour (\(H_{j}\)), as shown in Equation (2).

\begin{equation}\tag{2}
F_{i,H_{j}} = \sum_{t \in H_{j}}^{}U_{i,t}
\end{equation}

where \(t \in H_{j}\) refers to the four 15-minute intervals within hour \(H_{j}\).

Then, we could calculate the normalized average hourly frequency over \(D\) days (\({\overline{F}}_{i,H_{j}}^{norm}\)) as in Equation (3).

\begin{equation}\tag{3}
{\overline{F}}_{i,H_{j}}^{norm} = \frac{1}{D}\sum_{d = 1}^{D}\frac{F_{i,H_{j},d}}{N}
\end{equation}

where \(N\) was the number of intervals per hour, which was four in our dataset.

As noted, we considered the averaged frequency within a certain timeframe to enhance the behavioral use interpretability, as organized in Equation (4).

\begin{equation}\tag{4}
F_{i}^{p} = \frac{1}{|H_{p}|}\sum_{H_{j} \in H_{p}}^{}{\overline{F}}_{i,H_{j}}^{norm}
\end{equation}

where \(p\) was the timeframe (e.g., on-peak time from 4:00 pm to 9:00 pm).

The flexibility was calculated through Equations (5) and (6).

\begin{equation}\tag{5}
\sigma_{i}^{p} = \sqrt{\frac{1}{\left| H_{p} \right| \bullet D}\sum_{d = 1}^{D}{\sum_{H_{j} \in H_{p}}^{}\left( F_{i,H_{j},d}^{norm} - {\overline{F}}_{i,H_{p}}^{norm} \right)^{2}}}
\end{equation}
\begin{equation}\tag{6}
\sigma_{i}^{norm} = \frac{\sigma_{i} - \sigma^{\min}}{\sigma^{\max} - \sigma^{\min}\ }
\end{equation}

where \(\sigma^{\max}\) was the maximum standard deviation and \(\sigma^{\min}\) was the minimum standard deviation within the house.

Lastly, the average and variance of power usage (Equations 7 and 8) were calculated.

\begin{equation}\tag{7}
P_{i}^{avg,p} = \frac{1}{|H_{p}| \bullet D}\sum_{d = 1}^{D}{\sum_{H_{j} \in H_{p}}^{}P_{i,H_{j},d}}
\end{equation}
\begin{equation}\tag{8}
P_{i}^{var,p} = \sqrt{\frac{1}{|H_{p}| \bullet D}\sum_{d = 1}^{D}{\sum_{H_{j} \in H_{p}}^{}\left( P_{i,H_{j},d} - P_{i}^{avg,p} \right)^{2}}}
\end{equation}

These proposed metrics were discussed with an external subject expert for third-party adjudication, and their validity was confirmed after a series of discussions. Ultimately, we could pinpoint behavioral adjustments that can potentially lead to energy savings using these metrics, which were treated as reference solutions in our study.

\subsubsection{Home Data Selection and Saving Strategies}

We had access to energy use and generation data from 30 homes in Austin, collected at 15-minute intervals at both circuit and appliance levels during July and August, a period when peak loads often occur, purchased from Pecan Street Inc. \cite{r42}. For this experiment, we chose the best case for our experiment based on the following criteria: (1) number and diversity of appliances used during the data collection period in a home and (2) complexity in identifying feasible behavioral changes for energy savings.

\begin{figure}[tbp]
\centering
\includegraphics[width=0.60\textwidth]{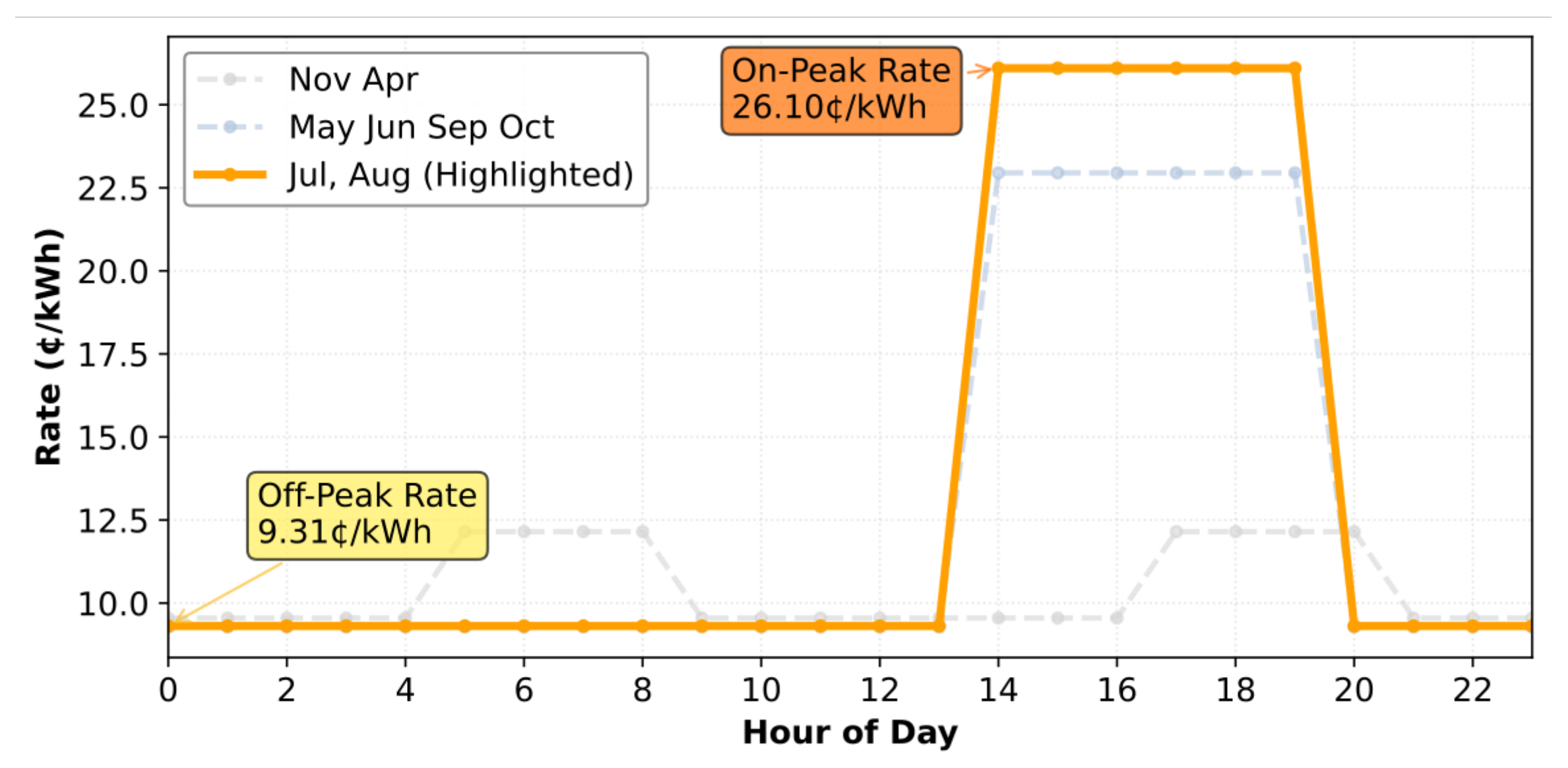}
\caption{TOU rate used in the role-playing experiment.}\label{fig:3}
\end{figure}

We obtained the actual TOU rate data from the utility rate database \cite{r43} and utilized it for our experimental study. Figure~\ref{fig:3} illustrates the TOU rate used in this study.

Table~\ref{tab:2} shows the analysis of the appliances and circuits used in the selected home using the proposed four metrics.

\begin{table}[tbp]
\centering
\caption{Appliance and circuit analysis of the selected house based on the proposed four metrics}\label{tab:2}
\includegraphics[width=0.87\textwidth]{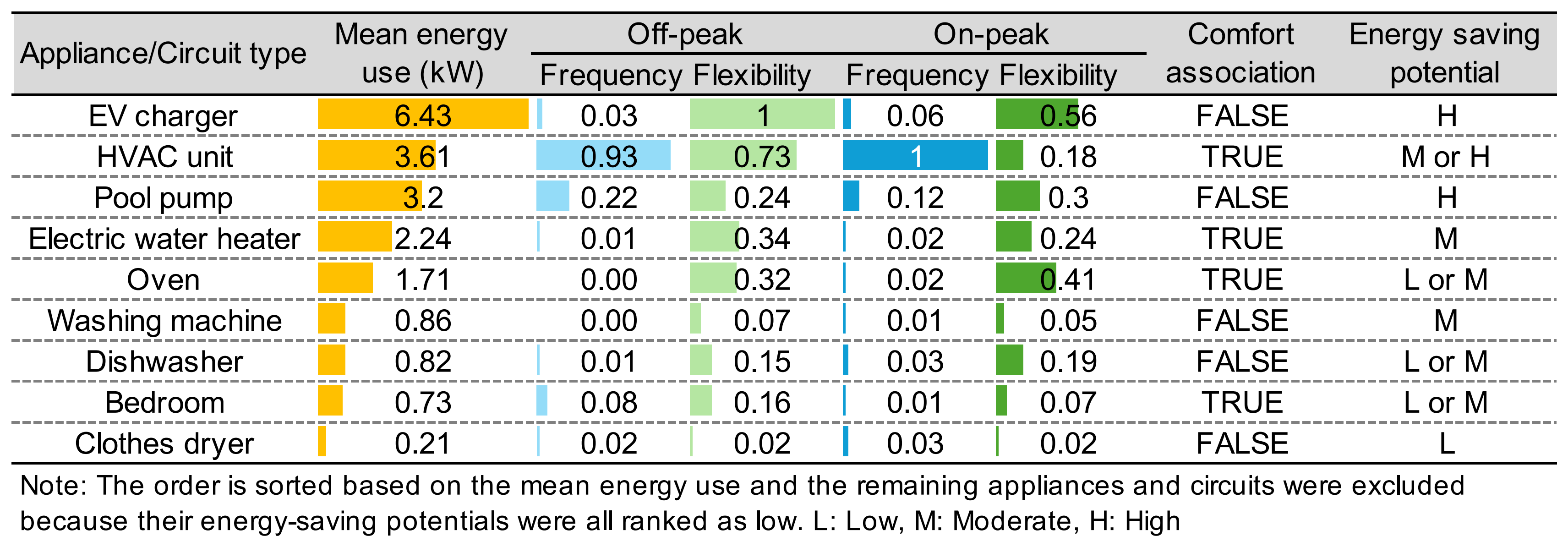}
\end{table}

Among those with high or moderate energy-saving potential (as indicated in the last column of Table~\ref{tab:2}), we selected top five appliances and formulated corresponding behavioral changes to achieve energy savings, as follows.

\begin{itemize}
\item
  \emph{HVAC unit}: This appliance was the most frequently used in this house and had the second highest mean energy consumption value (3.61kW) -- consequently, the highest energy driver of this house. Although it is closely associated with user comfort, several interventions can be viable, including:

  \begin{enumerate}
  \def\labelenumi{\alph{enumi}.}
  \item
    Co-conditioning with personal comfort systems (PCSs): As indicated in \cite{r44}, using PCSs -- which typically consume much less energy than HVAC units -- in conjunction with HVAC units such as portable electric fans can expand the range of temperature setpoint selection.
  \item
    Adjusting the HVAC unit's temperature setpoint: This intervention must be implemented carefully, as it does not guarantee user comfort. Hence, only a minimal temperature setback, e.g., 1-2°F or 0.6-1.1°C, should be considered to mitigate potential discomfort.
  \item
    Precooling strategy: If users are not sensitive to cooler temperatures during off-peak hours, they can precool their homes before on-peak hours.
  \end{enumerate}
\item
  \emph{Pool pump}: This appliance showed a high mean energy consumption (3.2kW) and was used more frequently in both on and off-peak hours, compared with other high energy-consuming appliances except for the HVAC unit. In addition, it showed moderate flexibility values in both on- and off-peak hours.
\item
  \emph{EV charger}: This appliance had the highest mean energy use (6.43kW) among all appliances and circuits in this house. Furthermore, it was used more frequently during on-peak hours and had high flexibility in both on- and off-peak hours. Lastly, this appliance is not associated with user comfort.
\item
  \emph{Dishwasher}: This appliance showed a moderate energy consumption (0.82kW) and flexibility, its energy saving potential was rated as moderate and low as it was not associated with user comfort.
\item
  \emph{Electric water heater}: This appliance showed a high mean energy consumption (2.24kW) and used more frequently during on-peak hours with moderate flexibility values in both timeframes. Despite its promise, its energy saving potential was assessed as moderate or low given its association with user comfort.
\end{itemize}

These possible behavioral changes were set as our answers to the role-playing experiment.

\subsection{Pre-Experiment: Subject Recruitment and Grouping}

This study advertised the human subject experiment, starting from Feb, 2025 to May, 2025 through emails, online advertisement (UA@Work), and personal invitations using authors' professional networks. Interested individuals were invited to complete a prescreening survey (provided in Appendix A) designed to assess their domain knowledge in home energy use and AI literacy levels for participant grouping.

The prescreening survey measured two constructs through a 5-point Likert scale: Domain knowledge was assessed through questions that captured participants' experiential and self-evaluated understanding of building energy, including their responsibility for utility bills, their familiarity with home energy use patterns. AI literacy was assessed through questions that captured both behavioral and self-evaluated aspects of GPT use such as the frequency and duration of GenAI tool use, self-evaluated expertise. These measures are self-reported, capturing participants' perceived rather than objective competence levels. However, in human-AI interaction contexts, perceived competence is arguably the more behaviorally relevant construct, as users' beliefs about their own knowledge and skills directly shape how they approach and engage with AI systems \cite{r45,r46}. This measurement approach is consistent with current practice in the field, where self-reported scales have been adopted to assess both domain knowledge \cite{r47,r48} and AI literacy \cite{r49,r50} in human-AI interaction research.

\begin{table}[tbp]
\centering
\caption{Four groups with different building energy knowledge and AI literacy levels}\label{tab:3}
\begin{tabular}{@{}llcc@{}}
\toprule
 & & \multicolumn{2}{c}{AI Literacy} \\
\cmidrule(l){3-4}
 & & Low & High \\
\midrule
\multirow{2}{*}{Domain Knowledge} & Low & Group LL & Group LH \\
 & High & Group HL & Group HH \\
\bottomrule
\end{tabular}
\end{table}

Using a stratified sampling approach, we attempted to distinguish subjects into four groups (Table~\ref{tab:3}). The goal of our stratified sampling approach is to have a balanced distribution across all groups in this study.

\subsection{Experiment}

\subsubsection{Experimental Sequence}

\begin{figure}[tbp]
\centering
\includegraphics[width=1.00\textwidth]{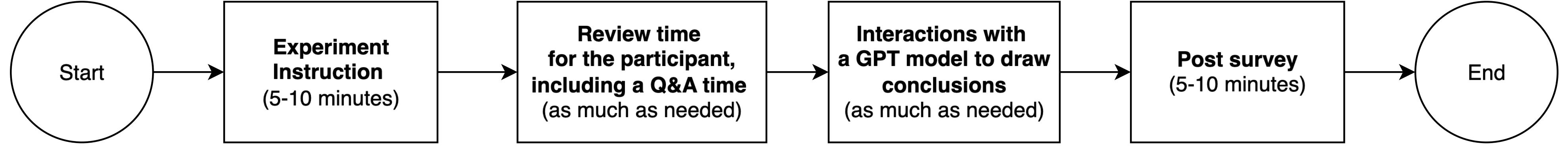}
\caption{Experimental sequence from the participant point of view.}\label{fig:4}
\end{figure}

Figure~\ref{fig:4} illustrates the sequence of the designed role-playing experiment. To enhance accessibility, we conducted the experiment using a laptop (in-person) or the videotelephony software program Zoom (online). Participants were first briefed on their goal -- reducing the household energy bill -- and given sufficient time to review and understand the details of the experiment provided through the instruction document (see Appendix for more details). Participants were allowed to raise any questions, and the experimenter clarified their questions. Next, participants were granted control of the ChatGPT 4o model in a web-browser (the selection of the GPT model is provided in Appendix) as illustrated in Figure~\ref{fig:5}. At the beginning of the role-playing experiment, two datasets were uploaded in the prompt. The objective was that the GPT model could process and analyze data based on participants' needs as if the GPT model was connected to energy data collected by appliance-level metering technologies. This step marked the temporary realization of an LLM-integrated BEMS, simulating how such a system could function to support data-driven energy analysis and decision-making. Then the experiment commenced. Participants were given limitless time to interact with the GPT tool to discuss and formulate their conclusions. When finished, the experimenter ensured that the chat remained within the OpenAI account and saved it for analysis. Lastly, participants were asked to take the post-experiment survey.

\begin{figure}[tbp]
\centering
\includegraphics[width=0.71\textwidth]{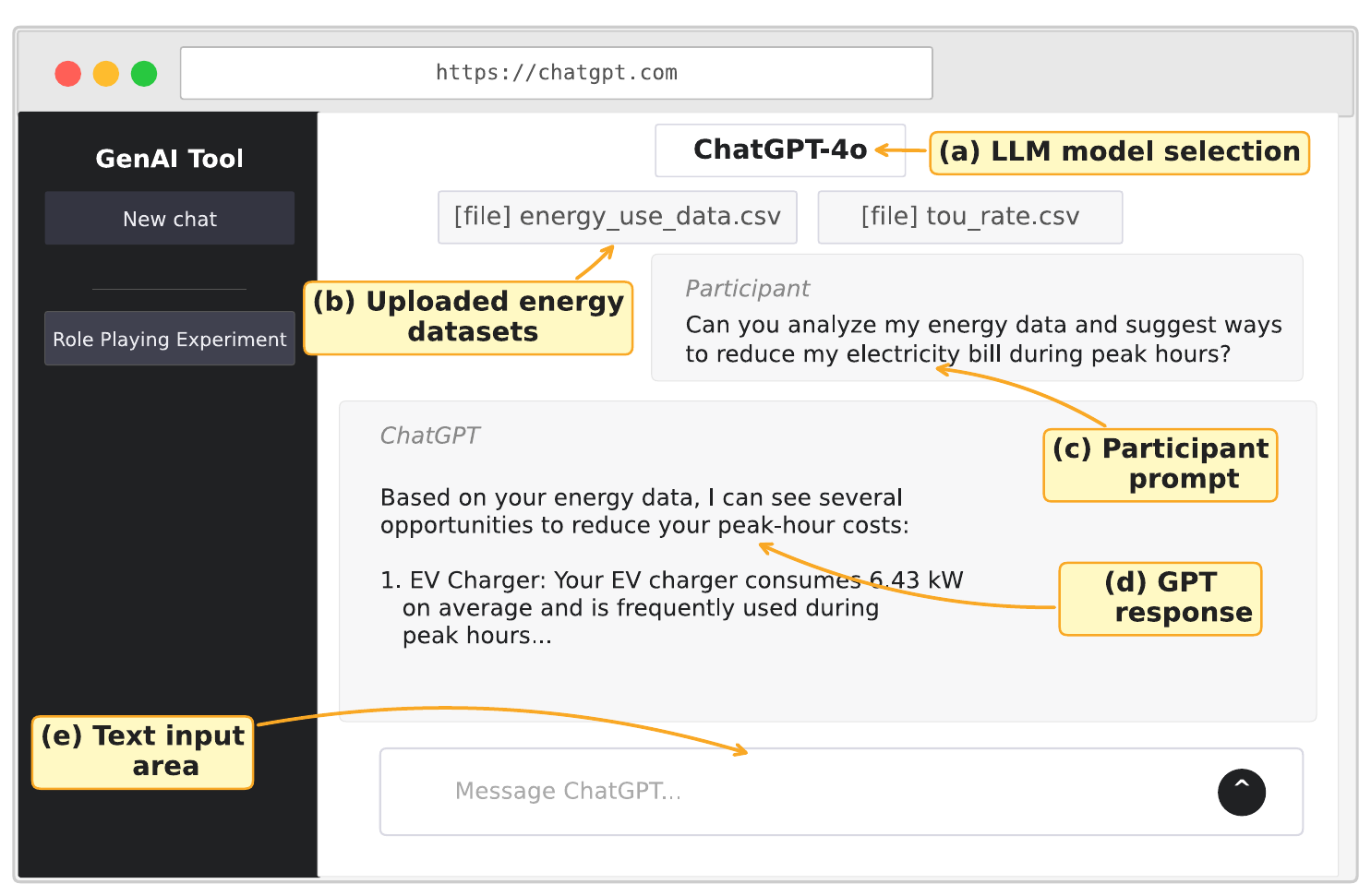}
\caption{Schematic of generative AI interface in the role-playing experiment}\label{fig:5}
\end{figure}

\subsubsection{Participant Information}

\begin{figure}[tbp]
\centering
\includegraphics[width=0.55\textwidth]{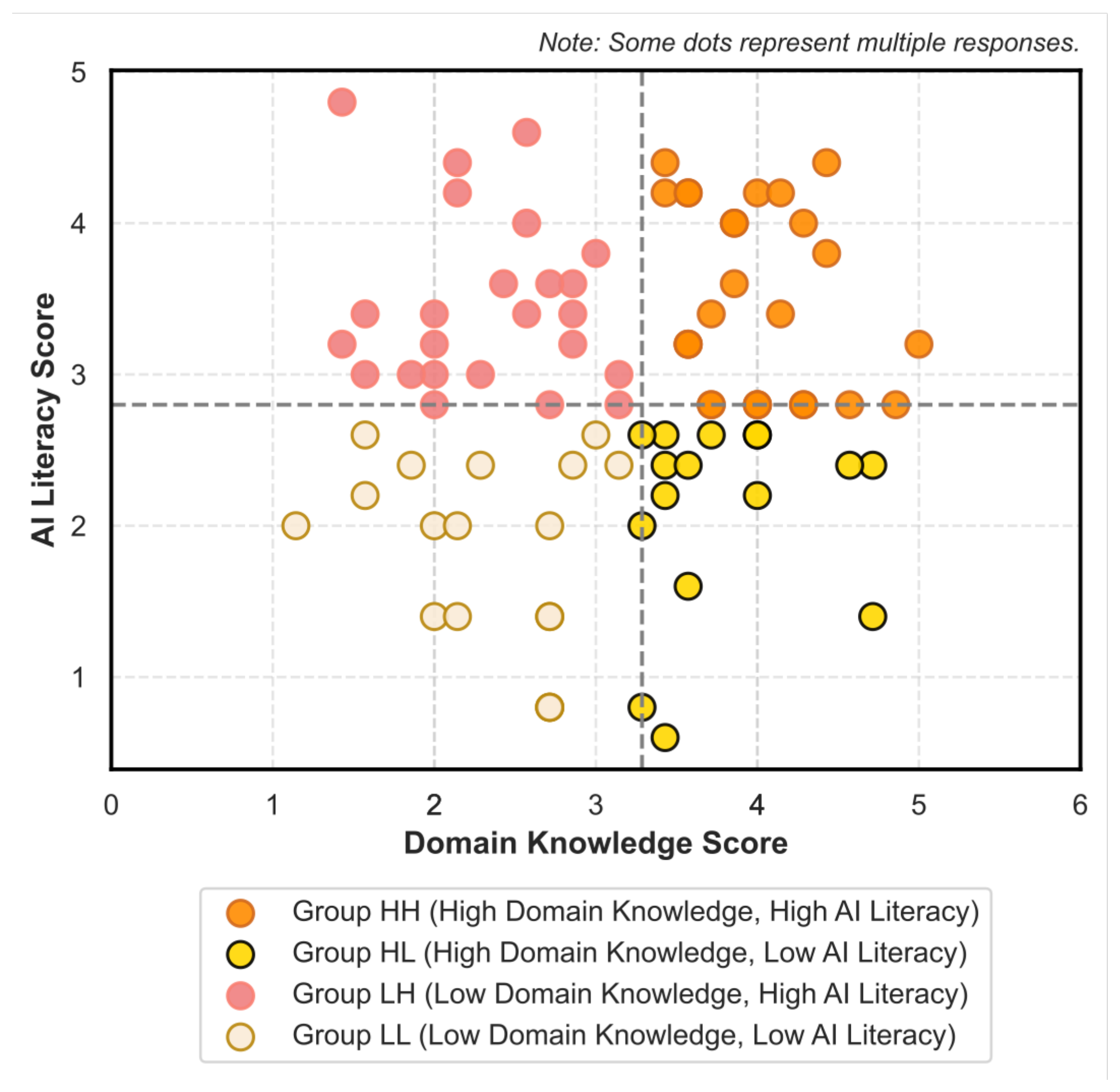}
\caption{Distribution of participants' domain knowledge scores and AI literacy scores.}\label{fig:6}
\end{figure}

Eligible participants were invited via email, and 85 individuals ultimately participated in the role-playing experiment out of the 114 who completed the pre-screening survey. Figure~\ref{fig:6} together illustrate the diversity and balance of the study's participant pool across two key dimensions: domain knowledge (building energy) and AI literacy. We classified participants into four groups based on median splits, revealing meaningful representation in each quadrant. This balanced distribution ensured that the study captured a full spectrum of user profiles, supporting robust comparisons and insights into how different types of users would engage with LLM-integrated BEMS. We provided participants' demographic characteristics in Appendix.

\subsection{Post-Experient: Prompt-Response Data Analysis}

\subsubsection{Overview}

\begin{figure}[tbp]
\centering
\includegraphics[width=1.00\textwidth]{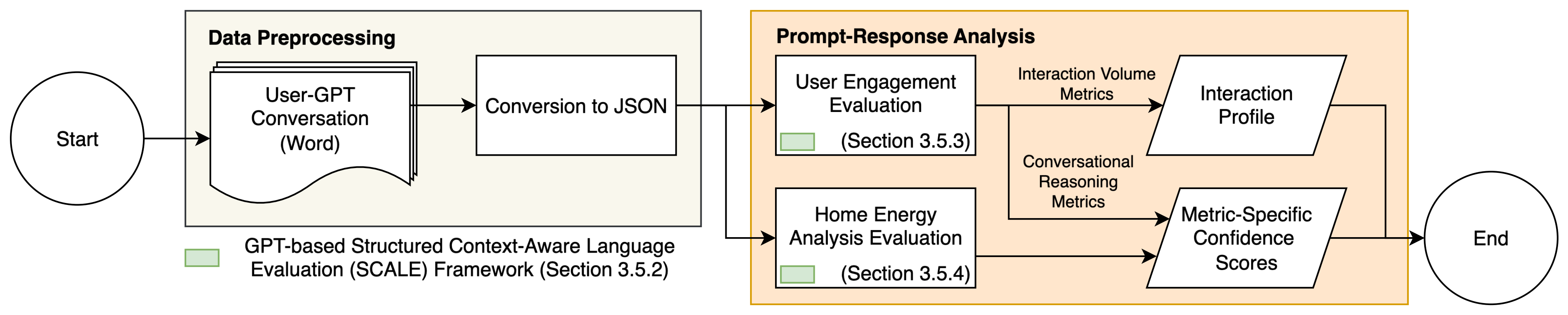}
\caption{Prompt-response data analysis flowchart}\label{fig:7}
\end{figure}

The prompt-response data were processed and analyzed through a systematic framework as shown in Figure~\ref{fig:7}. The goal was to establish a consistent and replicable scoring process that that could be applied across the collected dialogue data to meaningfully capture and represent the human-AI interaction dynamics.

\subsubsection{GPT-based Structured Context-Aware Language Evaluation (SCALE)}

\begin{figure}[tbp]
\centering
\includegraphics[width=1.00\textwidth]{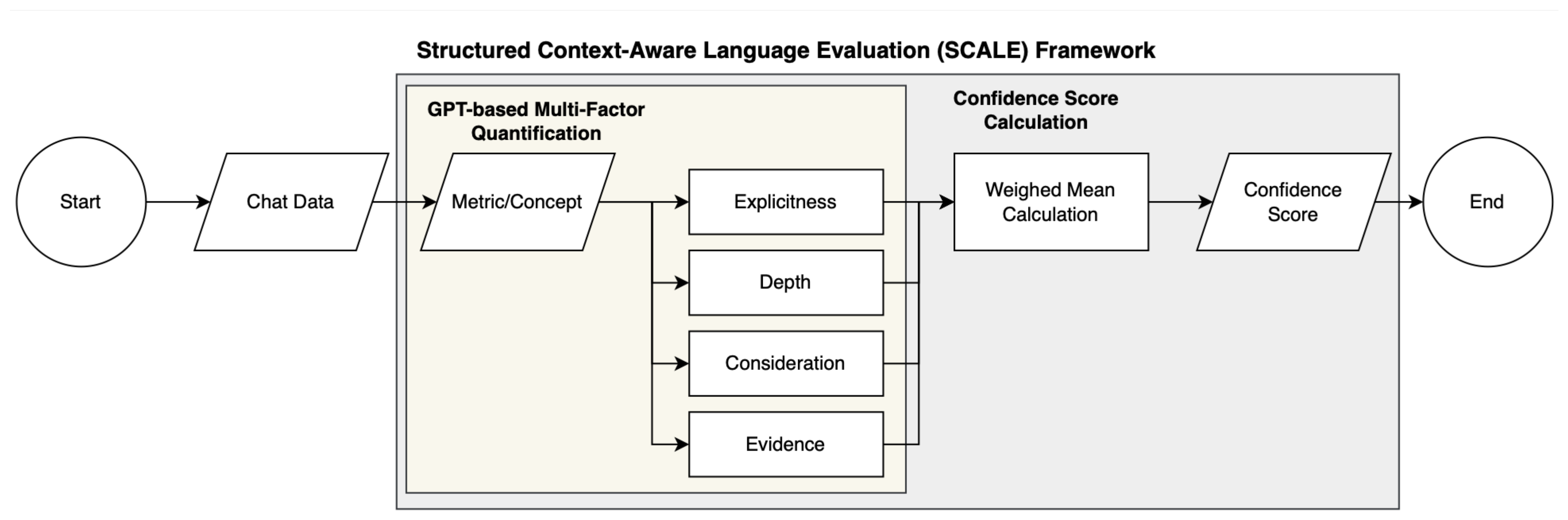}
\caption{The SCALE framework flowchart}\label{fig:8}
\end{figure}

The proposed GPT-based SCALE framework leveraged the contextual understanding capabilities of GPT, specifically, GPT-4o-mini as our analytical and scoring engine, to determine whether specific concepts or metrics are meaningfully addressed within a conversation as illustrated in Figure~\ref{fig:8}.

Rather than relying solely on keyword detection or pattern matching, this framework assesses the contextual relevance and clarity of user-GPT exchanges to classify whether a particular analytical element was actively considered. SCALE framework addressed a primary challenge in conversation analysis by offering a more accurate method for identifying substantive content.

The framework's multi-factor structure offered explainable confidence scores via four weighted factors with detailed justifications, as detailed in below:

\begin{itemize}
\item
  \textbf{Explicitness}: This factor measured whether participants used appropriate domain terminology and demonstrate clear understanding of concepts. It drew on linguistic research examining directness and clarity in communication \cite{r51}, and communication theory emphasizing the importance of explicit information exchange in human-computer interaction contexts \cite{r52}.
\item
  \textbf{Depth}: This factor evaluated whether participants engaged at a surface semantic level (e.g., basic acknowledgement, simple responses without elaboration) or a substantive latent level (e.g., detailed reasoning, comprehensive understanding, thoughtful analysis with context). It built on cognitive processing research distinguishing between shallow and deep levels of information engagement \cite{r53,r54} and thematic analysis research that examines different levels of analytical engagement with data \cite{r55}.
\item
  \textbf{Consideration}: Drawing on consumer behavior research on consideration sets \cite{r56}, this factor measured whether specific concepts or alternatives were actively incorporated into participants' decision-making process, reflecting the breadth of their evaluative reasoning.
\item
  \textbf{Evidence}: This factor emphasized the importance of supporting information in credible communication, building on discourse analysis research \cite{r57} that examines how speakers construct persuasive and well-substantiated arguments through the use of supporting details, examples, and justifications.
\end{itemize}

Then, the confidence score is calculated using Equation (1).

\begin{equation}\tag{1}
\text{Confidence score} = \frac{\omega_{1} \bullet \mathit{Explicitness} + \omega_{2} \bullet \mathit{Depth} + \omega_{3} \bullet \mathit{Consideration} + \omega_{4} \bullet \mathit{Evidence}}{4}
\end{equation}

\noindent Where \(\omega\) is the weight for each scoring factor. In this study, we applied the same weights to evenly consider all factors.

This framework was implemented systematically through a chain-of-thought prompt template with scoring rubrics for reproducibility. These details are provided in Appendix.

\subsubsection{User Engagement Evaluation}

The objective of this user engagement analysis was to assess the extent and style of participants' interactions with the GPT model. Consequently, seven metrics were employed as follows:

\begin{enumerate}
\def\labelenumi{\arabic{enumi}.}
\item
  \textbf{Interaction Volume}
\end{enumerate}

The proposed three metrics in this category captured key aspects of user interaction volume.

\begin{itemize}
\item
  \textbf{Total conversation turns}: This metric captured the total number of conversation turns -- comprising both user prompts and GPT responses -- required for a participant to reach their conclusion. The underlying premise is that fewer turns may indicate a more efficient interaction, though it may suggest greater reliance on GPT's analytical capabilities.
\item
  \textbf{Average prompt length}: This metric measured the average number of words per user prompt. It helps assess how much information or context participants provided to the GPT model throughout the conversation. Longer prompts may indicate more detailed or specific inquiries, reflecting greater user initiative or domain knowledge, while shorter prompts may suggest reliance on GPT to guide the interaction or formulate solutions.
\item
  \textbf{Prompt-response ratio}: This metric compared the number of user prompts to the number of GPT responses. It provides insights into the interactivity and balance of the conversation. A higher ratio may indicate more user-driven engagement and iterative querying, while a lower ratio may suggest that users relied more heavily on single, extended GPT responses or submitted fewer prompts overall.
\end{itemize}

\begin{enumerate}
\def\labelenumi{\arabic{enumi}.}
\setcounter{enumi}{1}
\item
  \textbf{Conversational Reasoning}
\end{enumerate}

We defined four distinct reasoning metrics, building on decision support systems theory \cite{r58}, which explores how interactive computer systems can enhance human decision-making processes through structured information gathering, alternative evaluation, and choice implementation. These metrics represented how participants interacted with GPT to find energy-saving interventions. The SCALE framework was applied for these metrics.

\begin{itemize}
\item
  \textbf{Information seeking}: This metric captured directive speech acts aimed at obtaining information, clarification, or elaboration. This metric was crucial for understanding how participants actively acquired knowledge and clarified concepts, which serves as a foundation for more informed energy decisions. Examples were: Can you explain that in more detail? how does this work? or what are the specific steps to implement this?
\item
  \textbf{Constraint articulation}: This metric aimed to evaluate assertive speech acts that state personal limitations, boundaries, or situation constraints. Capturing this behavior revealed how well users personalize solutions and set realistic boundaries, reflecting their ability to contextualize recommendations. Examples were: I am sensitive to being hot, that is infeasible for my situation, or I do not want to change my mealtime.
\item
  \textbf{Solution evaluation}: This metric was to identify assertive speech acts that express judgments, assessments, or evaluations of proposed solutions. This metric highlighted the participant's ability to critically assess and engage with the AI-generated suggestions, a key indicator of cognitive involvement in decision-making. Examples were: The proposed method seems reasonable, I think the oven idea makes sense, electric vehicle charging at night is an excellent idea, or this would be effective for my situation.
\item
  \textbf{Commitment expression}: This metric was to measure commissive speech acts that express willingness to try solutions or commit to behavioral changes. This metric was essential for assessing whether the interaction led to an actionable behavioral commitment or confirmation of recommended behavioral changes, serving as a proxy for the user's readiness or willingness to follow through with energy-saving actions suggested during the conversation. Examples were: I am willing to adjust my thermostat settings, I can reschedule when I run my dry washer, I will try using ceiling fans to comfort myself, or I will implement this change gradually.
\end{itemize}

\subsubsection{Home Energy Analysis Evaluation}

In the home energy analysis, we investigated how both the participants and the GPT model interpreted and reasoned through the provided home energy data to arrive at actionable conclusions. The objective was to understand how users engaged with the data by applying their building energy knowledge and utilizing the model's capabilities to identify feasible behavioral changes, reflecting a combination of human insight and AI support in the decision-making process.

In addition to energy-saving potential metrics introduced in Section 3.2, we incorporated additional three metrics to better capture participants' reasoning processes and understanding of home energy analysis as presented in Table~\ref{tab:4}. Collectively, these metrics allowed us to assess how participants integrated practical considerations when forming their strategies. The SCALE framework was applied in this analysis.

\begin{table}[tbp]
\centering\small
\caption{Evaluation metrics used for home energy analysis and how they were identified in SCALE}\label{tab:4}
\begin{tabularx}{\textwidth}{@{}>{\raggedright\arraybackslash}p{2.6cm}>{\raggedright\arraybackslash}X>{\raggedright\arraybackslash}X@{}}
\toprule
Metric & Definition & Examples \\
\midrule
Cost awareness & Discussion of electricity bills, energy costs, utility rates, TOU pricing, cost savings, or financial impact of energy decisions &
Your electricity bill was \$150 last month,\newline
Peak hours are more expensive,\newline
I want to reduce my energy bills,\newline
Please consider TOU rates in your calculations, \\
\addlinespace
Behavioral change & Discussion of adjusting usage patterns, adopting energy-saving behaviors, modifying habits, or actionable steps to reduce energy consumption &
I suggest running the dishwasher at night,\newline
Your HVAC running times can be adjusted to off-peak hours,\newline
Let's evaluate how I can change my energy usage patterns to reduce energy bills,\newline
I can't change my energy use patterns for the refrigerator, \\
\addlinespace
Technical knowledge & Discussion showing understanding of how appliances work, system operations, equipment specifications, or technical aspects of energy use &
Heat pumps are more efficient than traditional HVAC systems like furnaces,\newline
Smart thermostats can optimize energy use through scheduling and sensors,\newline
Variable-speed pumps adjust their speed based on demand, \\
\bottomrule
\end{tabularx}
\end{table}

\subsection{Post-Experiment: User-GPT Conclusion Analysis}

\begin{figure}[tbp]
\centering
\includegraphics[width=0.92\textwidth]{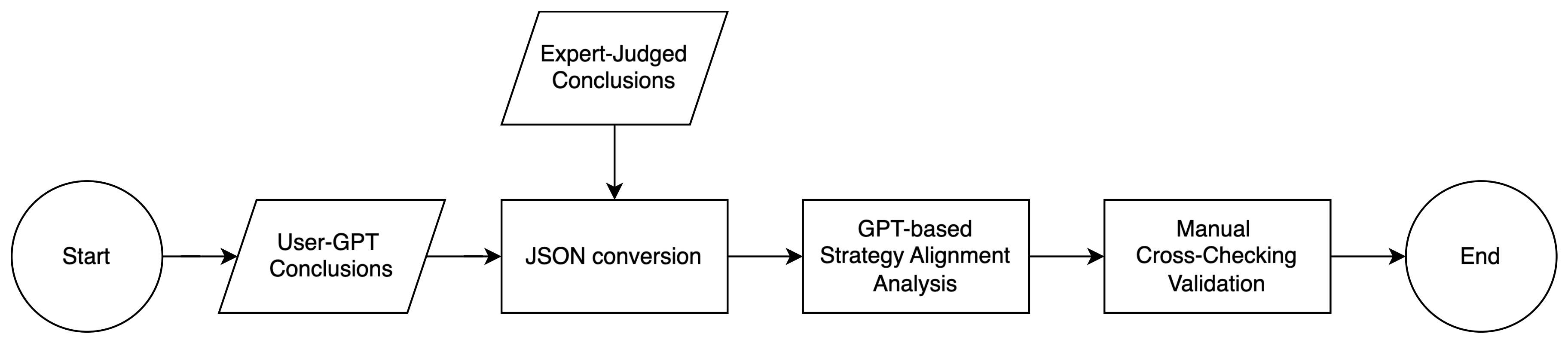}
\caption{User-GPT conclusion analysis using reference conclusions and GPT}\label{fig:9}
\end{figure}

To evaluate whether participants, through their interaction with the AI system, arrived at conclusions with reference solutions, a two-stage evaluation, informed by the LLM-as-a-judge paradigm \cite{r59}, was conducted as shown in Figure~\ref{fig:9}. In the first stage, the LLM (Open AI GPT-4o-mini) was used to systematically compare each participant's final energy-saving recommendations against a predefined set of reference solutions, following a CoT evaluation protocol that has been indicated to improve alignment with human judgements \cite{r60}. This comparison produced two distinct outcome metrics: (1) the appliance identification rate, defined as the proportion of the five target appliances that the participant correctly identified as candidates for energy savings, and (2) the strategy alignment rate, defined as the proportion of the seven reference strategies that the participant's recommendations substantially matched. In the second stage, the entire evaluation results from the LLM were manually cross-checked for three times as a validation step to ensure the reliability of the assessments.

\subsection{Post-Experiment: Survey Analysis}

The post-experiment survey, available in Appendix, was prepared: (1) to understand how participants would assess their own human-AI collaboration experiences (2) to assess changes in participants' building energy knowledge and experience with the GPT model and (3) to gather participants' perspectives on the benefits of utilizing GPT for managing home energy use. For the first goal, we posed questions about the role of their domain knowledge and prompt expertise during the role-playing experiment. For the second goal, we included a direct comparison between pre-screening and post-experiment responses by intentionally reusing the same questions from the prescreening survey -- listed below.

\begin{itemize}
\item
  Rate your understanding of home energy use and its patterns,
\item
  Rate your understanding of how monthly energy bills are calculated,
\item
  Rate your expertise in using GenAI tools like ChatGPT.
\end{itemize}

Regarding the third objective, we posed questions about whether they would consider using an AI-assisted BEM system in their homes and whether they believed such a system could support their learning and understanding of BEM. This line of questioning was intended to investigate both participants' openness to future adoption and their perception of GPT as a learning companion in the field of home energy decision-making.

\subsection{Individual, Group-Specific, and Overall Analysis and Comparison}

The analysis proceeded in two stages. First, we examined individual-level results across all participants to establish overall patterns, then conducted group-based comparisons to assess the influence of domain knowledge and AI literacy on interaction behaviors and outcomes. Four analytical dimensions were considered: user engagement levels, home energy analysis, comparison between user-GPT conclusions and reference solutions, and post-experiment survey responses. Descriptive statistics, including means, standard deviations, and medians, were computed for each metric across the four participant groups to characterize general trends. To formally test for group differences, we employed non-parametric inferential tests, selected due to the non-normal distributions observed across most metrics and the unequal group sizes resulting from the median-split classification. Kruskal-Wallis H tests \cite{r61} (degree of freedom = 3, $\alpha$ = .05) were first conducted to assess whether significant differences existed across the four groups. Mann-Whitney U tests \cite{r62} were then used to examine the main effects of each factor, domain knowledge (high vs. low) and AI literacy (high vs. low), independently, as well as all six pairwise group comparisons. Effect sizes were quantified using the rank-biserial correlation (r) \cite{r63}, and Bonferroni correction \cite{r64} was applied to pairwise comparisons to control for familywise error rate.

\section{Results}

\subsection{Case-based Analysis}

To illustrate how the analysis framework captures distinct patterns of user-GPT interaction, we presented three representative cases selected from 85 participants. These cases were selected to represent contrasting interaction patterns along two dimensions: user engagement level and GPT performance quality. As shown in Table~\ref{tab:5}, the scoring criteria differentiated these cases across multiple metrics, revealing how variations in user input quality and interaction depth correspond to differences in both SCALE confidence scores and alignment with reference solutions. The full conversation logs are provided in Appendix H.

\begin{table}[tbp]
\centering\small
\caption{Results derived from the analysis framework using three representative conversations}\label{tab:5}
\setlength{\tabcolsep}{4pt}
\begin{tabular}{@{}>{\raggedright\arraybackslash}p{2.7cm}>{\raggedright\arraybackslash}p{4.1cm}cccccc@{}}
\toprule
Category & Metric & \multicolumn{2}{c}{Example \#1} & \multicolumn{2}{c}{Example \#2} & \multicolumn{2}{c}{Example \#3} \\
\midrule
\multirow{3}{=}{Participant Information} & Group & \multicolumn{2}{c}{LL} & \multicolumn{2}{c}{LH} & \multicolumn{2}{c}{HH} \\
 & Domain knowledge & \multicolumn{2}{c}{2.71} & \multicolumn{2}{c}{2.57} & \multicolumn{2}{c}{4.43} \\
 & AI literacy & \multicolumn{2}{c}{1.25} & \multicolumn{2}{c}{3.75} & \multicolumn{2}{c}{3.75} \\
\addlinespace
\multirow{3}{=}{Interaction Volume} & Total Conversation Turns & \multicolumn{2}{c}{4.00} & \multicolumn{2}{c}{10} & \multicolumn{2}{c}{10} \\
 & Average Prompt Length & \multicolumn{2}{c}{8.50} & \multicolumn{2}{c}{17.00} & \multicolumn{2}{c}{54.6} \\
 & Prompt Response Ratio & \multicolumn{2}{c}{0.081} & \multicolumn{2}{c}{0.11} & \multicolumn{2}{c}{0.153} \\
\addlinespace
\multirow{4}{=}{Conversational Reasoning} & Information Seeking & \multicolumn{2}{c}{0.56} & \multicolumn{2}{c}{0.84} & \multicolumn{2}{c}{0.68} \\
 & Constraint Articulation & \multicolumn{2}{c}{0.00} & \multicolumn{2}{c}{0.00} & \multicolumn{2}{c}{0.68} \\
 & Solution Evaluation & \multicolumn{2}{c}{0.00} & \multicolumn{2}{c}{0.00} & \multicolumn{2}{c}{0.68} \\
 & Commitment Expression & \multicolumn{2}{c}{0.00} & \multicolumn{2}{c}{0.00} & \multicolumn{2}{c}{0.68} \\
\addlinespace
 & & User & GPT & User & GPT & User & GPT \\
\cmidrule(lr){3-4}\cmidrule(lr){5-6}\cmidrule(l){7-8}
\multirow{7}{=}{Home Energy Analysis} & Appliance energy use & 0.00 & 0.74 & 0.00 & 0.74 & 0.48 & 0.94 \\
 & Cost awareness & 0.56 & 0.74 & 0.66 & 0.74 & 0.52 & 0.94 \\
 & Behavioral change & 0.56 & 0.74 & 0.84 & 0.82 & 0.76 & 0.94 \\
 & Appliance use flexibility & 0.00 & 0.44 & 0.74 & 0.58 & 0.48 & 0.74 \\
 & Appliance use frequency & 0.00 & 0.00 & 0.00 & 0.52 & 0.00 & 0.62 \\
 & Comfort association with appliance & 0.00 & 0.44 & 0.00 & 0.00 & 0.66 & 0.70 \\
 & Technical knowledge & 0.00 & 0.56 & 0.00 & 0.74 & 0.00 & 0.70 \\
\addlinespace
User-GPT conclusion & Alignment rate with reference solutions & \multicolumn{2}{c}{80\%} & \multicolumn{2}{c}{20\%} & \multicolumn{2}{c}{100\%} \\
\bottomrule
\end{tabular}
\end{table}

\textbf{Example \#1.} This case showcases how a productive interaction can emerge even with minimal user engagement, provided GPT performs at a high level. As in Table~\ref{tab:5}, the interaction volume metrics reflected the participant's minimal engagement: the conversation consisted of only four total turns with a notably low prompt-response ratio, indicating that GPT generated substantially more content than the participant. The conversational reasoning metrics further confirmed this pattern. The participant's engagement was limited to information seeking in the initial prompt, with no constraint articulation, solution evaluation, or commitment expression observed throughout the conversation. This indicates that the participant neither refined the scope of the analysis nor critically assessed GPT's recommendations.

Despite the limited engagement, the home energy analysis metrics revealed a clear asymmetry: GPT consistently scored higher than the participant across most dimensions, particularly in appliance energy use and technical knowledge, where the participant made no substantive contributions. Nonetheless, the user-GPT conclusions achieved an 80\% alignment rate with reference solutions, with the only omission being a dishwasher intervention. This case demonstrates that GPT can produce reasonably accurate energy-saving interventions, even with limited user guidance, though the missed strategy suggests that more targeted prompting could have improved completeness.

\emph{\textbf{Example \#2}}. This case demonstrates how increased interaction volume does not necessarily lead to better outcomes when user engagement lacks critical evaluation of GPT's recommendations. In contrast to Example \#1, the interaction volume metrics indicated a substantially more active conversation: 10 total turns with an average prompt length of 17.00 words. However, the conversational reasoning metrics revealed that this increased activity did not translate into deeper analytical engagement. While information seeking was relatively high (0.84), constraint articulation, solution evaluation, and commitment expression all remained at zero. In other words, the participant asked multiple follow-up questions, requesting TOU rate schedules, per-appliance savings estimates, and annual projections, but never questioned the validity of GPT's initial recommendations or attempted to refine them. Each subsequent prompt sought additional clarification on the same set of strategies rather than evaluating whether those strategies were appropriate.

This lack of critical evaluation proved consequential. The home energy analysis metrics showed that GPT's initial response contained a mismatch between its data analysis and its recommendations, a pattern reflected in the low user-GPT alignment rate of only 20\% with reference solutions. Despite the participant's relatively high AI literacy, the low domain knowledge appears to have limited their ability to recognize this mismatch. This case illustrates that interaction volume alone is an insufficient indicator of productive human-AI collaboration; without substantive evaluation of AI-generated recommendations, additional turns may simply reinforce an initially suboptimal output.

\textbf{Example \#3}. This case showcases how high-quality user engagement can steer the conversation toward a fully aligned outcome, even when GPT's initial recommendations required refinement. Although the interaction volume was comparable to Example \#2 in total conversation turns (10), the nature of engagement differed substantially. The average prompt length was 54.6 words (more than three times that of Example \#2) and the prompt-response ratio was correspondingly higher. More critically, the conversational reasoning metrics distinguish this case from both preceding examples: all four dimensions were active throughout the conversation. This represents the only case among the three where the participant moved beyond information seeking to critically evaluate GPT's suggestions, articulate personal constraints, and propose alternative strategies.

This deeper engagement directly shaped the trajectory of the conversation. The participant rejected GPT's refrigerator optimization suggestion as infeasible, articulated thermal comfort constraints that precluded aggressive thermostat adjustments, proposed ceiling fans as an alternative cooling strategy, and assessed the feasibility of reducing pool pump runtime by questioning its impact on water quality. These interactions prompted GPT to revise its initial recommendations, ultimately producing two new interventions that aligned with expert-judged solutions. The home energy analysis metrics reflected this collaborative dynamic: the participant contributed meaningfully across most dimensions, particularly in behavioral change and comfort association, while GPT complemented these contributions with higher scores in appliance energy use, cost awareness, and technical knowledge. The resulting user-GPT conclusions achieved a 100\% alignment rate with expert-judged solutions. This case demonstrates that when participants actively evaluate, constrain, and redirect GPT's outputs, the collaborative process can yield outcomes that match expert-level recommendations.

\subsection{Overall \& Group-level Analysis}

This subsection presents the overall and group-level results across all 24 metrics, derived from the proposed analysis framework. Table~\ref{tab:6} summarizes the descriptive statistics

\begin{sidewaystable}
\centering\scriptsize
\caption{Summary of metric scores by participant group with Kruskal-Wallis test results}\label{tab:6}
\setlength{\tabcolsep}{2.2pt}
\begin{tabular}{@{}>{\raggedright\arraybackslash}p{2.4cm}>{\raggedright\arraybackslash}p{2.1cm}cccccccccc@{}}
\toprule
Category & Metric & \multicolumn{2}{c}{LL (n=28)} & \multicolumn{2}{c}{LH (n=24)} & \multicolumn{2}{c}{HL (n=12)} & \multicolumn{2}{c}{HH (n=21)} & \multicolumn{2}{c}{p} \\
\midrule
\multirow{3}{=}{Interaction Volume} & Total Conversation Turns & \multicolumn{2}{c}{13.21$\pm$14.20} & \multicolumn{2}{c}{12.88$\pm$10.40} & \multicolumn{2}{c}{13.42$\pm$11.96} & \multicolumn{2}{c}{10.90$\pm$8.05} & \multicolumn{2}{c}{0.94} \\
 & Average Prompt Length & \multicolumn{2}{c}{19.93$\pm$18.94} & \multicolumn{2}{c}{23.15$\pm$17.44} & \multicolumn{2}{c}{18.70$\pm$8.49} & \multicolumn{2}{c}{25.16$\pm$18.66} & \multicolumn{2}{c}{0.36} \\
 & Prompt Response Ratio & \multicolumn{2}{c}{0.10$\pm$0.07} & \multicolumn{2}{c}{0.14$\pm$0.12} & \multicolumn{2}{c}{0.14$\pm$0.12} & \multicolumn{2}{c}{0.14$\pm$0.11} & \multicolumn{2}{c}{0.790} \\
\addlinespace
\multirow{4}{=}{Conversational Reasoning} & Information Seeking & \multicolumn{2}{c}{0.70$\pm$0.12} & \multicolumn{2}{c}{0.67$\pm$0.18} & \multicolumn{2}{c}{0.68$\pm$0.08} & \multicolumn{2}{c}{0.69$\pm$0.10} & \multicolumn{2}{c}{0.80} \\
 & Constraint Articulation & \multicolumn{2}{c}{0.48$\pm$0.32} & \multicolumn{2}{c}{0.43$\pm$0.32} & \multicolumn{2}{c}{0.45$\pm$0.34} & \multicolumn{2}{c}{0.46$\pm$0.31} & \multicolumn{2}{c}{0.91} \\
 & Solution Evaluation & \multicolumn{2}{c}{0.41$\pm$0.32} & \multicolumn{2}{c}{0.47$\pm$0.30} & \multicolumn{2}{c}{0.39$\pm$0.30} & \multicolumn{2}{c}{0.40$\pm$0.33} & \multicolumn{2}{c}{0.85} \\
 & Commitment Expression & \multicolumn{2}{c}{0.29$\pm$0.35} & \multicolumn{2}{c}{0.25$\pm$0.35} & \multicolumn{2}{c}{0.25$\pm$0.32} & \multicolumn{2}{c}{0.16$\pm$0.27} & \multicolumn{2}{c}{0.58} \\
\addlinespace
 & & User & GPT & User & GPT & User & GPT & User & GPT & User & GPT \\
\cmidrule(lr){3-4}\cmidrule(lr){5-6}\cmidrule(lr){7-8}\cmidrule(lr){9-10}\cmidrule(l){11-12}
\multirow{7}{=}{Home Energy Analysis} & Appliance energy use & 0.25$\pm$0.31 & 0.76$\pm$0.07 & 0.34$\pm$0.30 & 0.76$\pm$0.06 & 0.25$\pm$0.32 & 0.73$\pm$0.06 & 0.29$\pm$0.29 & 0.77$\pm$0.09 & 0.79 & 0.85 \\
 & Cost awareness & 0.63$\pm$0.16 & 0.74$\pm$0.06 & 0.56$\pm$0.24 & 0.72$\pm$0.15 & 0.64$\pm$0.05 & 0.73$\pm$0.06 & 0.65$\pm$0.09 & 0.75$\pm$0.07 & 0.55 & 0.80 \\
 & Behavioral change & 0.67$\pm$0.17 & 0.77$\pm$0.06 & 0.67$\pm$0.17 & 0.75$\pm$0.06 & 0.66$\pm$0.10 & 0.75$\pm$0.06 & 0.67$\pm$0.09 & 0.75$\pm$0.11 & 0.60 & 0.45 \\
 & Appliance use flexibility & 0.20$\pm$0.28 & 0.54$\pm$0.17 & 0.34$\pm$0.30 & 0.55$\pm$0.13 & 0.20$\pm$0.30 & 0.54$\pm$0.19 & 0.22$\pm$0.27 & 0.49$\pm$0.22 & 0.27 & 0.58 \\
 & Appliance use frequency & 0.00$\pm$0.00 & 0.23$\pm$0.27 & 0.00$\pm$0.00 & 0.25$\pm$0.28 & 0.04$\pm$0.15 & 0.21$\pm$0.26 & 0.00$\pm$0.00 & 0.24$\pm$0.28 & 0.11 & 0.93 \\
 & Comfort association with appliance & 0.27$\pm$0.32 & 0.49$\pm$0.26 & 0.28$\pm$0.31 & 0.38$\pm$0.29 & 0.28$\pm$0.35 & 0.48$\pm$0.17 & 0.33$\pm$0.33 & 0.45$\pm$0.24 & 0.87 & 0.46 \\
 & Technical knowledge & 0.02$\pm$0.10 & 0.60$\pm$0.29 & 0.00$\pm$0.00 & 0.58$\pm$0.31 & 0.00$\pm$0.00 & 0.48$\pm$0.36 & 0.05$\pm$0.16 & 0.63$\pm$0.28 & 0.32 & 0.87 \\
\addlinespace
\multirow{3}{=}{User-GPT conclusion} & Overall alignment & \multicolumn{2}{c}{0.65$\pm$0.13} & \multicolumn{2}{c}{0.71$\pm$0.13} & \multicolumn{2}{c}{0.63$\pm$0.12} & \multicolumn{2}{c}{0.67$\pm$0.13} & \multicolumn{2}{c}{0.22} \\
 & Appliance identification rate & \multicolumn{2}{c}{0.69$\pm$0.17} & \multicolumn{2}{c}{0.77$\pm$0.17} & \multicolumn{2}{c}{0.63$\pm$0.14} & \multicolumn{2}{c}{0.76$\pm$0.17} & \multicolumn{2}{c}{\textbf{0.04\textsuperscript{*}}} \\
 & Strategy alignment rate & \multicolumn{2}{c}{0.59$\pm$0.17} & \multicolumn{2}{c}{0.62$\pm$0.17} & \multicolumn{2}{c}{0.52$\pm$0.15} & \multicolumn{2}{c}{0.57$\pm$0.12} & \multicolumn{2}{c}{0.17} \\
\bottomrule
\multicolumn{12}{@{}l@{}}{Note: Mean $\pm$ standard deviation; \textsuperscript{*}statistically significant} \\
\end{tabular}
\end{sidewaystable}

\textbf{Interaction Volume.} Across all four groups, participants engaged in a comparable number of conversation turns with the AI system, averaging approximately 12.6 turns per session. Neither domain knowledge nor AI literacy produced meaningful differences in how much participants interacted with the system. Similarly, average prompt length and the prompt-response ratio, reflecting the verbosity and relative contribution of user input, did not differ across groups. These results suggest that participants across all competence profiles adopted broadly similar interaction patterns in terms of volume and balance of exchange, regardless of their prior expertise.

\textbf{Conversational Reasoning.} The four SCALE-based conversational reasoning dimensions showed no significant group differences. Information seeking scores were relatively uniform across groups, indicating that all participants, regardless of background, directed the conversation toward gathering relevant energy-related information at similar rates. Constraint articulation and solution evaluation scores were moderate across the board but exhibited high within-group variability, suggesting that individual differences in how participants framed constraints or critically assessed GPT-generated solutions far exceeded any group-level tendencies. Commitment expression remained the lowest-scoring dimension overall, reflecting that most participants did not explicitly commit to specific strategies during the conversation, a pattern consistent across all groups.

\textbf{Home Energy Analysis.} On the user side, participants demonstrated moderate engagement with cost awareness and behavioral change, while energy consumption understanding remained relatively low with high variability. Use frequency and technical knowledge were near floor levels, indicating that participants rarely discussed appliance usage patterns or technical details in their prompts. None of the seven user-side dimensions differed significantly across groups.

On the GPT side, ChatGPT consistently produced responses with high scores on energy consumption, cost awareness, and behavioral change, regardless of which group the participant belonged to. The uniformity of GPT-side scores across groups confirms that GPT's analytical output was largely driven by its own response generation tendencies rather than being shaped by differences in user input. Notably, the gap between user- and GPT-side scores was most pronounced for energy consumption and technical knowledge, suggesting that ChatGPT compensated for areas where participants contributed the least.

\textbf{User-GPT Conclusion Analysis.} The conclusion analysis evaluated whether participants' final energy-saving recommendations aligned with expert-validated reference solutions. The overall alignment score did not differ significantly across groups. However, the appliance identification rate was the only metric across all 24 measures to reach statistical significance. Groups with low AI literacy identified target appliances at higher rates than groups with high AI literacy. This counterintuitive pattern suggests that participants with greater AI experience may have relied more heavily on GPT's direction without independently verifying appliance-level conclusions, whereas less AI-literate participants may have drawn more on their own reasoning to confirm or supplement GPT's recommendations. The strategy alignment rate showed a similar directional trend but did not reach significance.

Post-experiment survey results, which capture participants' subjective perceptions of the LLM-integrated BEMS are presented in Section 4.3 alongside the corresponding hypothesis evaluations.

\subsection{Hypothesis Evaluation}

In this subsection, we assessed our hypotheses, presented in Introduction.

\begin{enumerate}
\def\labelenumi{\arabic{enumi}.}
\item
  \emph{Participants with limited domain knowledge may rely heavily on the LLM's capabilities, potentially resulting in brief interactions.}

  \textbf{Supported.} The tendency of relying on LLM's analytical capabilities to draw their conclusions was found across all participant groups regardless of their domain knowledge levels. Total conversation turns did not differ significantly across groups (H = 0.42, p = .936), nor did average prompt length (p = .358) or prompt-response ratio (p = .790). Participants with low domain knowledge (Groups LH and LL) averaged 12.88 and 13.21 turns, respectively, comparable to or slightly above the high domain knowledge groups (HH: 10.90, HL: 13.42). Conversational reasoning scores were similarly uniform, with no SCALE dimension showing significant group differences (all p \textgreater{} .57).

  This may indicate that participants, irrespective of domain expertise, viewed the LLM as a sufficiently efficient problem-solver \cite{r47,r65} -- reducing their perceived need to engage in deeper, iterative dialogue or to interrogate the model's outputs critically or to interrogate the model's outputs critically. Rather than validating or expanding on GPT's suggestions, many treated the GPT model more as a static information provider than a dynamic thought partner. This behavior suggested a broader cognitive framing of LLMs as answer engines rather than collaborative reasoning tools, especially when task goals seem achievable through direct prompting. It may also reflect a possible overestimation of GPT's analytical capabilities \cite{r66}, where participants, even those with high domain knowledge, defaulted to passive use rather than applying their expertise to guide, refine, or challenge the GPT model's responses. It is also worth noting that participants did not have direct access to the raw energy data during the conversation --- the datasets were uploaded to ChatGPT, which performed all numerical analysis. Without independent tools to verify GPT's calculations, participants across all groups were structurally positioned to rely on the model's analytical outputs, which may partially explain the uniform reliance pattern observed.

  This behavioral pattern underscored a critical challenge in human-AI collaboration: effective use of LLMs requires not only prompt engineering skills or domain knowledge in both users and GPT, but also a shift in user mindset -- from passive querying to active co-construction \cite{r67}.
\item
  \emph{Participants with strong domain knowledge may engage in extended interactions with the tool to validate LLM's suggestions, ultimately arriving at reasonably accurate conclusions.}

  \textbf{Not Supported}. Interestingly, participants with high domain knowledge did not consistently engage in longer or more complex interactions -- no statistical difference in the interaction volume or conversational reasoning metrics -- despite their presumed capacity to critically evaluate and refine the LLM's outputs, compared to their less knowledgeable counterparts. This counterintuitive result can be explained as follows: First, participants with strong domain expertise might have assumed that GPT was already equipped with sufficient knowledge, leading them to trust its responses without feeling the need to probe further -- a pattern consistent with automation complacency . Second, these participants may have approached the task with confidence in their own judgment, using GPT primarily as a confirmation tool rather than as a collaborator \cite{r68}. In such cases, their goal may not have been to challenge or extend the model's analysis, but rather to verify that it aligned with their existing understanding \cite{r69}. Moreover, high-knowledge participants might have found GPT's responses to be ``good enough'' within the constraints of the task, reducing their motivation to test alternative perspectives or explore edge cases \cite{r70}.

  This aligns with the observed pattern of relatively short interactions even among expert users, and challenges the assumption that domain expertise necessarily translates into deeper or more interactive use of AI tools.
\item
  \emph{Participants paired with strong AI literacy would compensate their limited domain knowledge by effectively interacting with GPT to extract relevant insights, formulate well-structured prompts, and critically assess AI-generated recommendations.}

  \textbf{Partially Supported.} AI literacy was the only participant factor that produced a statistically significant effect on any outcome metric (Table~\ref{tab:6}). Notably, Group LH achieved the highest alignment rate (77\%) among all four groups, outperforming groups with high domain knowledge. This suggests that strong AI literacy can partially offset limited technical expertise by enabling users to better engage with, interpret, and refine GPT's suggestions.

  However, this compensatory effect did not manifest through the specific mechanisms predicted by H3. None of the interaction volume or conversational reasoning metrics differed significantly by AI literacy. Therefore, it indicates that high AI literacy participants did not formulate notably longer prompts, engage in more conversation turns, or demonstrate measurably higher levels of constraint articulation or solution evaluation. This compensation appears to have operated not through how much or how explicitly participants interacted with GPT, but rather through how effectively they interpreted and acted upon GPT's outputs \cite{r67,r71}.

  We therefore consider that this hypothesis is partially supported, since a 77\% appliance identification rate still falls short of ideal collaboration outcomes. If Group HH had successfully collaborated with GPT in this role-playing experiment -- actively leveraging their domain knowledge to shape, question, and refine GPT's outputs -- we could have better observed the full extent to which the knowledge gap could be compensated through user-GPT collaboration. The fact that Group HH did not outperform LH reinforces the interpretation from H2: domain expertise alone did not translate into deeper analytical engagement with the LLM.
\item
  \emph{Excessively verbose GPT responses may negatively affect user agency and limit iterative dialogue, especially in participants with lower AI literacy.}

  \textbf{Supported.} The prompt-response ratio across all 85 participants was notably low (median: 0.08) and this indicates that GPT responses were roughly 12 times longer than participant prompts. This asymmetry suggests that the sheer volume of GPT-generated content might have overwhelmed participants -- making the interaction feel more like passively receiving information from AI than collaboratively working toward optimized solutions \cite{r72,r73}. The density and breadth of responses may have imposed a cognitive load that interfered with participants' ability to engage in critical reasoning or iterative inquiry, ultimately reducing the likelihood of deeper collaboration and shared decision-making regardless of their domain knowledge. This tendency was observed across all four groups. In other words, the structure and delivery of the AI's output to user prompts are critical factors that shape user engagement.

  This finding carries important design implications for LLM-based BEMS. Effective structure and delivery of LLM outputs should encourage clarity, accessibility, and opportunities for users to respond, reflect, and refine their energy-saving strategies in collaboration with the AI. One potential approach is offering adaptive output modes, which allow users to choose between detailed, comprehensive explanations and more concise, modular responses that invite follow-up questions. Such configurability would accommodate diverse use preferences while preserving opportunities for the iterative dialogue.
\item
  \emph{Participants would perceive this LLM-based analysis platform for HEM as useful and educational, providing a valuable opportunity to enhance their understanding while exploring feasible energy-saving strategies.}

  \textbf{Supported.} The post-experiment survey results strongly support this prediction. Six survey items captured participants' subjective evaluations of their interaction with GPT during the role-playing experiment, organized into two categories: self-evaluation of the experience and self-assessed human-AI collaboration (Figure~\ref{fig:10}).
\end{enumerate}

\begin{figure}[tbp]
\centering
\includegraphics[width=0.95\textwidth]{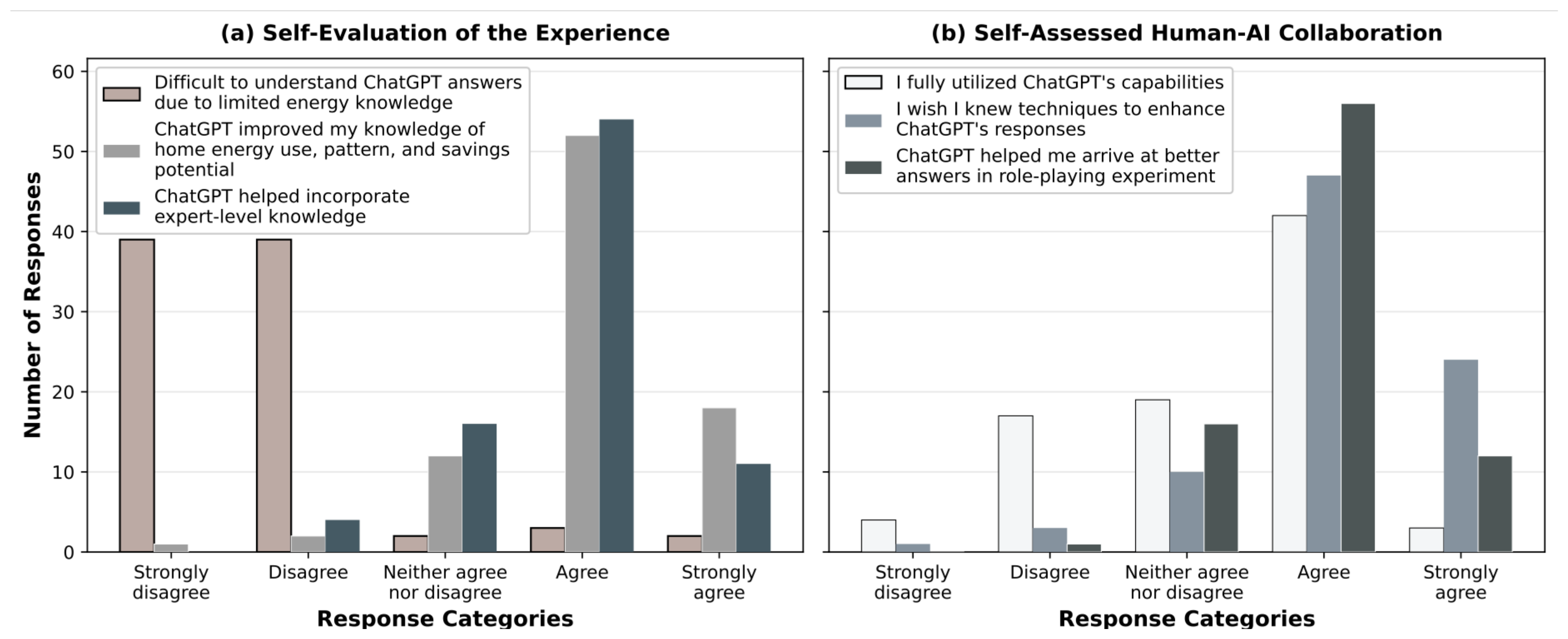}
\caption{Self-assessed human-AI collaboration}\label{fig:10}
\end{figure}

Regarding self-evaluation of the experience, responses to ``difficulty in understanding ChatGPT answers'' were overwhelmingly positive across all participant groups, with the majority disagreeing with the statement. Only five participants (5.9\%) chose either agree or strongly agree with this statement. Responses to the second question (improving knowledge of home energy use, patterns, and saving potential) revealed strong perceived learning benefits. For example, 91.7\% of participants in Group LH, the highest rate across all participant groups, either agreed or strongly agreed with the statement, and 71.4\% of Group HH shared the same view. Participant perceptions of GPT's capability to deliver expert-level support were similarly favorable. Interestingly, 81.8\% of participants with high building energy knowledge either agreed or strongly agreed with this statement while only 73.1\% of participants with low building energy knowledge shared the same view. This gap suggests that participants with higher domain knowledge might have had clearer expectations or benchmarks for expert-level reasoning, making them more receptive to GPT's ability to deliver technical insight and more confident in recognizing its expert-like contributions \cite{r74}.

Regarding self-assessed human-AI collaboration, responses to the first option (I fully utilized ChatGPT's capabilities) showed a moderate level of agreement across all groups. Agreement rates ranged from 39.3\% in Group LL to 70.8\% in Group LH, suggesting that participants with higher AI literacy felt more capable of leveraging the tool's full potential. The ``I wish I knew some techniques to enhance ChatGPT's responses'' question revealed a strong, shared desire for better prompt engineering knowledge even by the high AI literacy group, who 77.8\% expressed this sentiment. Group LL showed the highest agreement (96.5\%), indicating a widespread recognition, especially among novice users, of the gap between knowing how to use GPT and knowing how to employ it effectively. The last question (ChatGPT helped me arrive at better answers in the role-playing experiment) revealed an interesting trend across participant groups. The lowest agreement rate of 67.9\% was found from Group LL while the highest agreement rate of 85.7\% was found from Group HH.

\begin{figure}[tbp]
\centering
\includegraphics[width=1.00\textwidth]{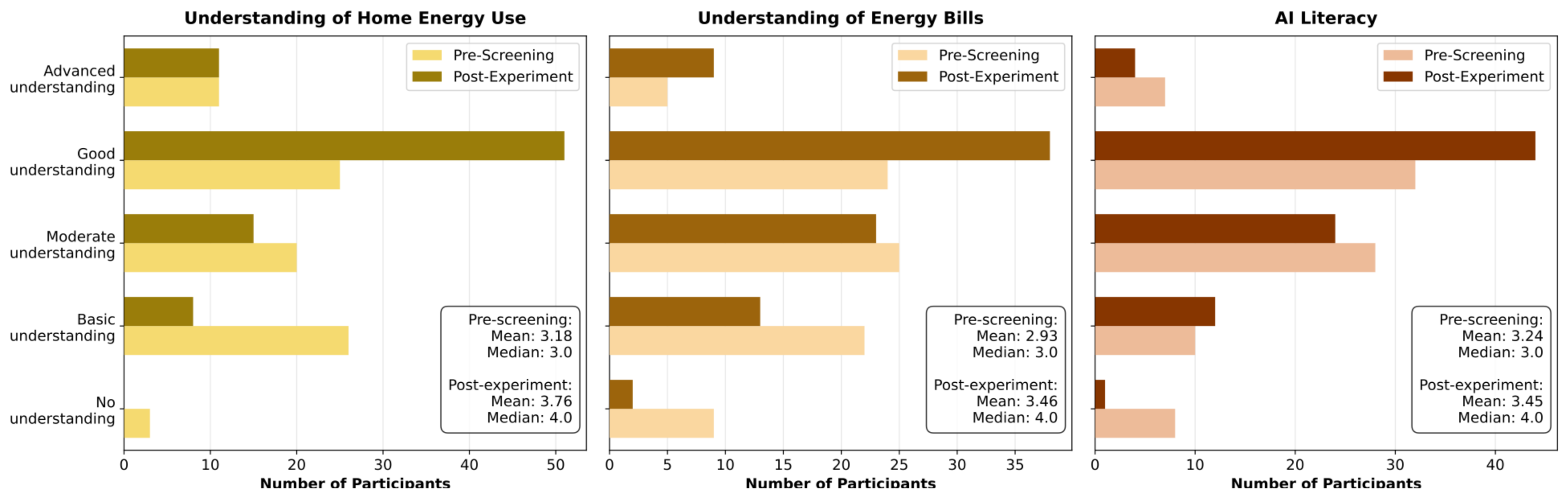}
\caption{Overall trends of self-assessed domain knowledge and GenAI tool expertise before and after the role-playing experiment.}\label{fig:11}
\end{figure}

These findings underscore the promise of LLM-based BEM platforms as dual-purpose tools, capable of supporting decision-making while also fostering energy literacy. This perception is further corroborated by the pre-post comparison of participants' self-reported assessments (Figure~\ref{fig:11}), which demonstrates a consistent upward shift in understanding of home energy use, understanding of energy bills, and AI literacy following the role-playing experiment. Taken together, these results suggest that the envisioned platform was not only perceived as a helpful assistant for immediate task completion but also served as a resource for ongoing learning and empowerment in navigating the increasingly complex energy landscape at home.

\section{Discussion}

This research, addressing core research questions in the interdisciplinary field of integrating AI with BEMS, left several discussion points.

\subsection*{Behavioral Tendency in User Prompts}

As analyzed through user engagement interaction volume metrics, most participants tended to utilize simple prompts to initiate their interactions with GPT. This brevity reflects a broader behavioral pattern where users might have expected GPT to deliver complete, accurate, and comprehensive answers without asking for detailed or iterative inputs -- potentially handing over the responsibility of providing the context to GPT. This behavioral tendency suggests that participants may have assumed GPT would infer or fill in contextual gaps on their behalf, which ultimately constrained the depth and interactivity of the conversation. Ultimately, this prompt approach consequently limited deeper exploration or iterative reasoning, suggesting that even prompt-savvy users might default to passive engagement when the task feels goal-oriented or when early responses from GPT seem satisfactory. Notably, this observed behavior may reflect how many people engage in real-world scenarios, asking a single, task-oriented question and accepting the first plausible response without further probing, thereby the interactive and co-creative potential of human-AI collaboration \cite{r75}.

\subsection*{Key Factor in Human-AI Collaboration: Domain Knowledge}

Recent research has demonstrated that while prompt engineering skills can significantly enhance the quality of LLM outputs, domain knowledge, incorporated into the LLM model, plays a critical role in formulating meaningful prompts and evaluating responses \cite{r76}. Currently, various techniques are available in LLMs to enhance performance on specific topics or applications such as parameter-efficient fine-tuning, RAG, and agentic frameworks. Even though these techniques could enhance LLM's analytical performance on home energy use data, we intentionally chose not to implement them in this study in order to concentrate more on observing the baseline capabilities of general-purpose LLMs when interacting with participants in a naturalistic, un-augmented setting. This decision allowed us to better isolate and assess the impact of user characteristics -- particularly, domain knowledge and AI literacy -- on the success of user-GPT collaboration, without confounding the results with system-level enhancements.

Previous studies also indicated that domain experts who acquire basic prompting skills achieve better results than AI-guided prompt engineering without domain knowledge \cite{r77}. Aligning with this observation that users require sufficient domain knowledge for high performance, this study discovered that in a human-AI collaboration context, users must also engage with AI at a deeper cognitive level -- by critically interpreting AI-generated outputs, questioning assumptions, and iteratively refining conclusions through reasoning. This level of engagement enables the collaboration to move beyond surface-level exchanges and toward more accurate, context-aware, personalized, and reference outcomes.

\subsection*{AI Literacy as the Differentiating Factor in Outcome Quality}

While much of the existing literature on human-AI collaboration has emphasized domain knowledge and prompt engineering skills as primary determinants of effective LLM use, the findings of this study suggest that AI literacy may also play a consequential role in shaping collaboration outcomes. Among the 20 metrics examined, the only statistically significant group difference was in the appliance identification rate, and this difference was driven by AI literacy.

This distinction carries important implications. If AI literacy primarily enhances how users process and act upon AI outputs rather than how they formulate inputs, then interventions aimed at improving human-AI collaboration should focus not only on teaching users to write better prompts but also on developing their capacity for critical evaluation of AI-generated content. Tankelevitch et al. \cite{r67} have similarly argued that generative AI imposes unique metacognitive demands on users. Specifically, it requires them to assess the relevance, completeness, and accuracy of outputs that may appear fluent and authoritative regardless of their actual quality. The present findings lend empirical support to this argument: the gap between high- and low-AI-literacy participants manifested not in their ability to extract information from GPT, but in their ability to discern what to do with it.

\subsection*{Role of GPT's Analytical Performance in Human-AI Collaboration}

GPT's analytical performance was not a targeted variable under investigation in this research; however, it emerged as a critical factor influencing the quality of human-AI collaboration, especially in shaping user-GPT conclusions. The majority of participants, even those with 100\% alignment rates, relied heavily on GPT's (vanilla) analytical capabilities for interpreting appliance-level home energy data rather than offering explicit guidance or specifying tailored computational approaches. Consequently, GPT's surface-level analysis was often sufficient to identify the most salient energy-saving strategies, such as those involving HVAC and pool pump usage (each with \textgreater94\% success rates), but fell short of uncovering less obvious, high-impact opportunities. In other words, if an LLM (or advanced LLM framework) shows strong contextual reasoning and analytical capabilities tailored to BEMS, it could enhance the productivity and effectiveness of human-AI collaborations by supporting deeper, more personalized, and data-driven exchanges. Our decision to use a general-purpose LLM in our experiment helped reveal the limitations of relying solely on default model behavior as our key takeaway.

\subsection*{Design Implications for LLM-Integrated BEMS}

The behavioral patterns and outcome differences observed in this study point toward several design considerations for LLM-integrated BEMSs. First, the pervasive passivity observed across all participant groups, and it suggests that users may benefit from scaffolded interaction designs. Rather than presenting a blank text input field and expecting users to formulate effective queries independently, platforms could incorporate guided prompting mechanisms such as structured question templates, suggested follow-up queries, or progressive disclosure interfaces that encourage users to refine their initial requests \cite{r78,r79}. Such scaffolding could be particularly valuable for users with lower domain knowledge, who may lack the vocabulary or conceptual framing to articulate specific analytical needs. Second, the finding that AI literacy was the key differentiator in outcome quality suggests that platforms should invest in interpretive support features. These might include confidence indicators for AI-generated recommendations, side-by-side comparisons of alternative strategies, or summary dashboards that help users evaluate and prioritize AI suggestions rather than passively accepting them \cite{r33,r80}. Such features would lower the metacognitive barrier for users with less AI experience, effectively embedding some of the evaluative capacity that high-AI-literacy participants brought to the task organically. Finally, given that GPT's surface-level analysis was sufficient for identifying salient energy-saving strategies but fell short on less obvious opportunities, platforms targeting professional or advanced use cases should consider incorporating domain-specific analytical layers \cite{r28,r81}. While the present study intentionally used baseline GPT to isolate user-level factors, real-world deployments would benefit from augmenting the LLM's analytical depth to better match the expectations of domain-knowledgeable users.

\section{Conclusion}

This study investigated how user domain knowledge and AI literacy shape human-AI interactions in an LLM-integrated BEMS. Using a 2×2 factorial design with 85 participants interacting with a GPT model to anlayze residential energy data, we assessed both the process of interaction through interaction volume and conversational reasoning metrics based on the SCALE framework and the quality of outcomes, measured by appliance identification and strategy alignment accuracy.

The key contributions and findings of this research are as follows:

\begin{itemize}
\item
  \textbf{LLM-assisted analysis produced remarkably uniform interaction patterns across expertise levels.} Across 19 process-level metrics spanning interaction volume and conversational reasoning, no statistically significant differences were found among the four participant groups. Neither domain knowledge nor AI literacy significantly altered how participants interacted with GPT, including conversation length, prompt complexity, or the types of reasoning expressed. This suggests that LLM tools may act as equalizers, flattening expertise-based differences in how users engage with analytical tasks. While this points to the accessibility and democratizing potential of LLM-integrated platforms, it also raises concerns about underutilization of domain expertise in human-AI collaboration.
\item
  \textbf{AI literacy emerged as the key differentiator in outcome quality.} The only statistically significant group difference was in the appliance identification rate (Kruskal-Wallis H = 8.49, p = .037), and this difference was driven by AI literacy (Mann-Whitney U, p = .009) rather than domain knowledge (p = .446). Participants with higher AI literacy accomplished more accurate conclusions, suggesting that AI literacy operates primarily at the stage of output interpretation and critical evaluation rather than input formulation.
\item
  \textbf{The SCALE framework was introduced as a structured analytical tool} for identifying and categorizing specific conversational concepts within human-AI interactions. Its five dimensions offer a systematic lens for assessing user reasoning quality. Its adaptable architecture supports broader application in future studies involving dialog-based AI systems.
\item
  \textbf{Two complementary metric sets were proposed} to assess user prompts and engagement levels. Interaction volume metrics captured the quantitative aspects of dialogue, while conversational reasoning metrics outlined user reasoning depth and interaction quality. Together, these metrics enabled a multi-layer assessment of how participants engaged with GPT to achieve their intended outcomes.
\item
  \textbf{Participants perceived the platform as useful and educational.} Post-experiment survey results indicated that the majority of participants found the LLM-assisted energy analysis valuable, and self-assessed knowledge levels in energy use and energy bills increased significantly from pre- to post-experiment, confirming the platform's educational potential even in a single-session interaction.
\end{itemize}

Beyond its research contributions, this study represents a foundational step toward realizing effective and scalable LLM-integrated BEMS that can empower a wide range of users to make informed, personalized, and context-sensitive energy decisions in their daily lives. Importantly, such systems can offer great opportunities for households, but also for utilities and policymakers. For utilities, where energy consumption and generation data are collected at scale, LLM-integrated BEMSs can support synthesizing energy use patterns at granular levels and facilitate the development of new and more customized DR programs -- realizing more effective peak demand management. Additionally, by benefitting from LLMs' ability to interpret trends and patterns in historical and real-time data, utilities could better segment customer groups, tailor communications for maximum program participation, and offer education opportunities. For policymakers, these systems offer a powerful means to promote energy literacy and evaluate the impact of policy interventions on energy use by analyzing dialogues between the systems and users -- specifically, how users interpret, engage with, and respond to the interventions. Such insights can help policymakers identify gaps and refine programs to be more inclusive and effective across diverse populations.

We acknowledge the limitations of this study. First, the experiment was designed for participants to act as if they were the household members making energy-saving decisions for the assigned home, as outlined in the experiment instruction. However, they were not the actual residents of the home and therefore lacked firsthand experience with its energy use patterns and behavioral routines. This gap may have limited their ability to question, reflect upon, or refine during their exchanges with GPT, potentially leading to reduced interaction volume and fewer opportunities for reasoning. Second, this the experimental design provided a single interaction session, which may have restricted participants' ability to experiment with prompt variations or build more advanced interactions over time. This one-shot format may have discouraged exploratory prompting and reinforced the use of short, directive inputs, meaning that the observed passivity in user behavior may partly reflect the constraints of the study protocol rather than participants' inherent interaction tendencies. Third, domain knowledge and AI literacy were assessed through self-reported survey items; hence, the inherent subjectivity of this evaluation approach should be acknowledged. Also, we divided them based on median values in two dimensions, which may have obscured differences among them. However, as noted above, perceived competence could be the more relevant construct in this human-AI interaction context as users' beliefs about their own capabilities directly influence their interaction behaviors with AI systems. Additionally, while standardized AI literacy instruments have been proposed in general educational contexts \cite{r50}, no validated scale currently exists that jointly captures domain knowledge in building energy management and AI literacy specific to LLM-based conversational tools. The survey items used in this study were therefore designed to reflect the constructs most relevant to this experimental context, and future work could benefit from developing and validating domain-specific measurement instruments. Fourth, this study relied on a single household energy dataset with a specific appliance configuration, climate context, and TOU rate structure. This design choice was intentional: providing all participants with identical data ensured that observed differences in interaction behaviors could be attributed to participant characteristics, especially domain knowledge and AI literacy, rather than variation in the underlying energy data. However, we acknowledge that the specific energy analysis outcomes, such as which appliances were identified and the nature of the energy-saving recommendations generated, may be influenced by the salience of particular appliances (e.g., HVAC, pool pump, EV charger) and the characteristics of the dataset used. Fifth, we employed ChatGPT-4o as our engine but given the rapid pace of advancements in this field, responses generated by more advanced LLMs may differ. Nonetheless, our core contribution remains valid, informative, and robust. One technical limitation worth noting was our use of JSON files in data processing. Finally, although JSON files facilitated structured data extraction and analysis, they prevented us from capturing and analyzing the visualizations that participants used to interpret the energy use data during the experiment. We decided not to overcomplicate our analysis by assessing these visual outputs as textual interactions could explain what was visualized. However, we acknowledge that incorporating visualization data could offer richer insights into how visual aids influence user reasoning, strategy development, and overall engagement with LLM.

To continue our contribution to realizing LLM-integrated BEMS, several directions for future research are planned. First, we will explore additional analysis of our dataset by incorporating participants' demographic information in the pre-screening survey (age, education level, gender, and ethnicity) to examine whether demographic heterogeneity contributes to differences in interactions within LLM-integrated BEMS. Second, we plan to investigate the potential of RAG or agentic frameworks in LLM-integrated BEMS, which can supplement LLMs with relevant external information (e.g., policy or rate updates) and can empower LLMs to autonomously plan and execute multistep tasks, guiding users through complex energy analyses. Third, future research can recruit actual homeowners to interact with the LLM-integrated BEMS about their own energy consumption over time to better simulate resident knowledge and promote intrinsic motivation, and address researchers to better capture the full range of human-AI interaction behaviors and assess whether sustained engagement leads to deeper collaboration and improved outcomes over time.

\acknowledgments{This material is based upon work supported by Salt River Project under grant SRP UA-08 24-25 and the National Science Foundation under grant \#2519054. We would like to appreciate our SRP advisor, Victor J. Berrios, who provided external evaluation of our methodological approach. In addition, we appreciate all participants to the experiment.}
\aiuse{During the preparation of this work the authors used ChatGPT 4o and Claude Sonnet 4 in order to proofread sentences and write Python codes to generate figures. After using these tools/services, the authors reviewed and edited the content as needed and take full responsibility for the content of the publication.}

\appendix
\setcounter{table}{0}\setcounter{figure}{0}
\renewcommand{\thetable}{A\arabic{table}}\renewcommand{\thefigure}{A\arabic{figure}}
\section*{Appendix}
\section{Pre-screening survey}

\begin{surveyblock}
This pre-screening survey is to understand your background information, knowledge about building energy, and experiences with large language model (LLM) tools like ChatGPT. We will collect your email address to schedule the actual experiment. We plan to conduct experiments using the stratified sampling approach based on the provided information.

\par\medskip\noindent\textbf{Background}\par

Q1: What is your age range?

\begin{enumerate}
\def\labelenumi{\arabic{enumi})}
\item
  18-24
\item
  25-29
\item
  30-34
\item
  35-39
\item
  40-44
\item
  45-49
\item
  50-54
\item
  55-59
\item
  60 and above
\end{enumerate}

Q2: Which gender do you identify yourself?

\begin{enumerate}
\def\labelenumi{\arabic{enumi})}
\item
  Male
\item
  Female
\item
  Non-binary / third gender
\item
  Prefer not to say
\end{enumerate}

Q3: Are you of Hispanic, LatinX, or Spanish origin?

\begin{enumerate}
\def\labelenumi{\arabic{enumi})}
\item
  No
\item
  Yes
\end{enumerate}

Q4: How would you describe your ethnicity?

\begin{enumerate}
\def\labelenumi{\arabic{enumi})}
\item
  White
\item
  Black or African American
\item
  American Indian or Alaska Native
\item
  Asian
\item
  Native Hawaiian or Pacific Islander
\item
  From multiple races
\item
  Some other races (please specify) \_\_\_\_\_\_\_\_\_\_\_\_\_\_\_\_\_\_\_\_
\item
  Prefer not to disclose
\end{enumerate}

Q5: What is the highest level of education you have completed or the highest degree you have received?

\begin{enumerate}
\def\labelenumi{\arabic{enumi})}
\item
  Less than high school degree
\item
  High school degree or equivalent
\item
  Some college but no degree
\item
  Associate degree
\item
  Bachelor degree
\item
  Master degree
\item
  Doctoral degree
\end{enumerate}

\par\medskip\noindent\textbf{Understanding of Building Energy}\par

Q6: Are you responsible for paying the monthly utility bills (e.g., electricity, gas, water) of your residence?

\begin{enumerate}
\def\labelenumi{\arabic{enumi})}
\item
  Yes
\item
  No
\end{enumerate}

Q7: Please rate your understanding of home energy use and its patterns

\begin{enumerate}
\def\labelenumi{\arabic{enumi})}
\item
  No understanding
\item
  Basic understanding
\item
  Moderate understanding
\item
  Good understanding
\item
  Complete understanding
\end{enumerate}

Q8: Please rate your understanding of a building system and appliance from the perspective of energy consumption.

\begin{center}\footnotesize
\begin{tabular}{@{}>{\raggedright\arraybackslash}p{4.6cm}*{5}{>{\centering\arraybackslash}p{1.75cm}}@{}}
\toprule
 & No understanding & Basic understanding & Moderate understanding & Good understanding & Complete understanding \\
\midrule
HVAC systems & & & & & \\
Lighting systems & & & & & \\
Building envelope (e.g., (walls, windows, roof, floors) & & & & & \\
Appliances (e.g., computers, monitors, refrigerators, etc.) & & & & & \\
\bottomrule
\end{tabular}
\end{center}

Q8: Are you familiar with time-of-use rates?

\begin{enumerate}
\def\labelenumi{\arabic{enumi})}
\item
  Definitely not
\item
  Probably not
\item
  Probably yes
\item
  Definitely yes
\end{enumerate}

\par\medskip\noindent\textbf{Experience in GenAI}\par

Q9: How often do you use Generative Artificial Intelligence (GenAI) tools such as ChatGPT, Claude, or Gemini in a typical day?

\begin{enumerate}
\def\labelenumi{\arabic{enumi})}
\item
  Never
\item
  Rarely (a few times a week)
\item
  Sometimes (a few times a day)
\item
  Frequently (several times a day)
\item
  Constantly (multiple times an hour)
\end{enumerate}

Q10: How long have you been using these GenAI tools?

\begin{enumerate}
\def\labelenumi{\arabic{enumi})}
\item
  less than a month
\item
  1-3 months
\item
  4-6 months
\item
  7-12 months
\item
  1-2 year(s)
\item
  More than 2 years
\end{enumerate}

Q11: Please rate your level of expertise in using GenAI tools like ChatGPT.

\begin{enumerate}
\def\labelenumi{\arabic{enumi})}
\item
  Beginner
\item
  Lower intermediate
\item
  Intermediate
\item
  Upper intermediate
\item
  Expert
\end{enumerate}

Q12: Please rank your familiarity with prompt engineering

\begin{enumerate}
\def\labelenumi{\arabic{enumi})}
\item
  Not at all familiar
\item
  Slightly familiar
\item
  Somewhat familiar
\item
  Moderately familiar
\item
  Extremely familiar
\end{enumerate}
\end{surveyblock}

\section{Experiment Instruction}

\begin{surveyblock}
In this experiment, you will play a role-playing game.

\textbf{Role:}

\begin{itemize}
\item
  You are a homeowner in Austin, Texas (Hot and Humid) and currently collecting circuit-, appliance-level power use data (a 15-min interval) for two months (July and August). Below is the list of circuits and appliances that you have data.

  \begin{itemize}
  \item
    Appliances: HVAC unit, electric vehicle charger, washing machine, dishwasher, clothes dryer, microwave, oven 1, oven 2, pool pump, cooktop, refrigerator, and electric water heater
  \item
    Circuits: Bathroom, bedroom, living room, and kitchen.
  \end{itemize}
\item
  You are responsible for this house's utility bills and currently participating in a time-of-use (TOU) rate (more details below).

  \begin{itemize}
  \item
    Time-of-use (TOU) rate: an electricity pricing structure where the cost of electricity varies depending on the time of day you use it. The figure below shows your TOU rate during summer and what you are using in your house.
  \end{itemize}
\end{itemize}

(Figure~\ref{fig:3} is presented here)

\textbf{Goal:}

\begin{itemize}
\item
  Your \textbf{goal} is to find \textbf{the most effective five behavioral changes} to \underline{reduce your energy bills} through a conversation with ChatGPT. Specifically, please interact with ChatGPT to identify potential five behavior-based energy-saving strategies of your house using the collected data.
\end{itemize}

\textbf{Caveat \& Restriction}:

\begin{itemize}
\item
  You also own other appliances in your house (you can imagine your house), but their energy use was not measured and collected.
\item
  The house energy data reflects your normal energy consumption behavior.
\item
  The behavioral changes should not significantly compromise your comfort or normal routines (e.g., delaying meals by two hours from your regular schedule or disruptions to your thermal comfort).
\item
  Behavioral changes do not mean that you intend to adopt new technology.
\end{itemize}

When the experimenter gives the signal, you may begin your interaction with ChatGPT.

\textbf{Post Survey:}

\begin{itemize}
\item
  After the role-playing experiment, you will take the post survey using Qualtrics and this will be the last step for you!
\end{itemize}
\end{surveyblock}

\section{GPT Model Selection Process}

Among high-performance GPT models, in Table A2, available during the time when the experimental study was conducted. We selected the OpenAI GPT 4o model for our experiment due to its superior multimodal capabilities, processing efficiency, and improved contextual understanding compared to other versions. Given that participants needed to analyze energy consumption data and provide calculation-based insights dynamically, GPT-4o's reasoning ability using user-provided datasets made it the most suitable selection. Additionally, GPT-4o provided lower latency and rapid responsiveness that ensured seamless and smoother user experiences, especially in a real-time role-playing interaction based on screen-sharing or remote-control environment while minimizing the technical disruptions.

\begin{table}[tbp]
\centering\small
\caption{GPT models available during the experiment}\label{tab:A1}
\begin{tabularx}{\textwidth}{@{}>{\raggedright\arraybackslash}p{2.0cm}>{\raggedright\arraybackslash}p{2.1cm}>{\raggedright\arraybackslash}Xc@{}}
\toprule
Model & Release date & Description & Ref \\
\midrule
GPT 4o & May 13, 2024 & A flagship multimodal model optimized for complex reasoning, creativity, and high-performance capabilities. In particular, it excels in data analytics of user-provided data, providing real-time seamless interaction. & \cite{r82} \\
\addlinespace
GPT 4o-mini & Jul 18, 2024 & A lightweight and faster version of GPT-4o, designed for efficiency and cost-effectiveness. However, it has reduced capacity for highly complex reasoning tasks, and multimodal scalability. & \cite{r83} \\
\addlinespace
GPT o1-mini & Sep 12, 2024 & A compact and highly efficient variant of GPT-o1, ideal for quick responses and lightweight application. However, it is not optimized for deep reasoning, large context handling, and real-time analytical interaction. & \cite{r84} \\
\bottomrule
\end{tabularx}
\end{table}

\section{Experiment Participant Demographic Information}

\begin{figure}[tbp]
\centering
\includegraphics[width=0.80\textwidth]{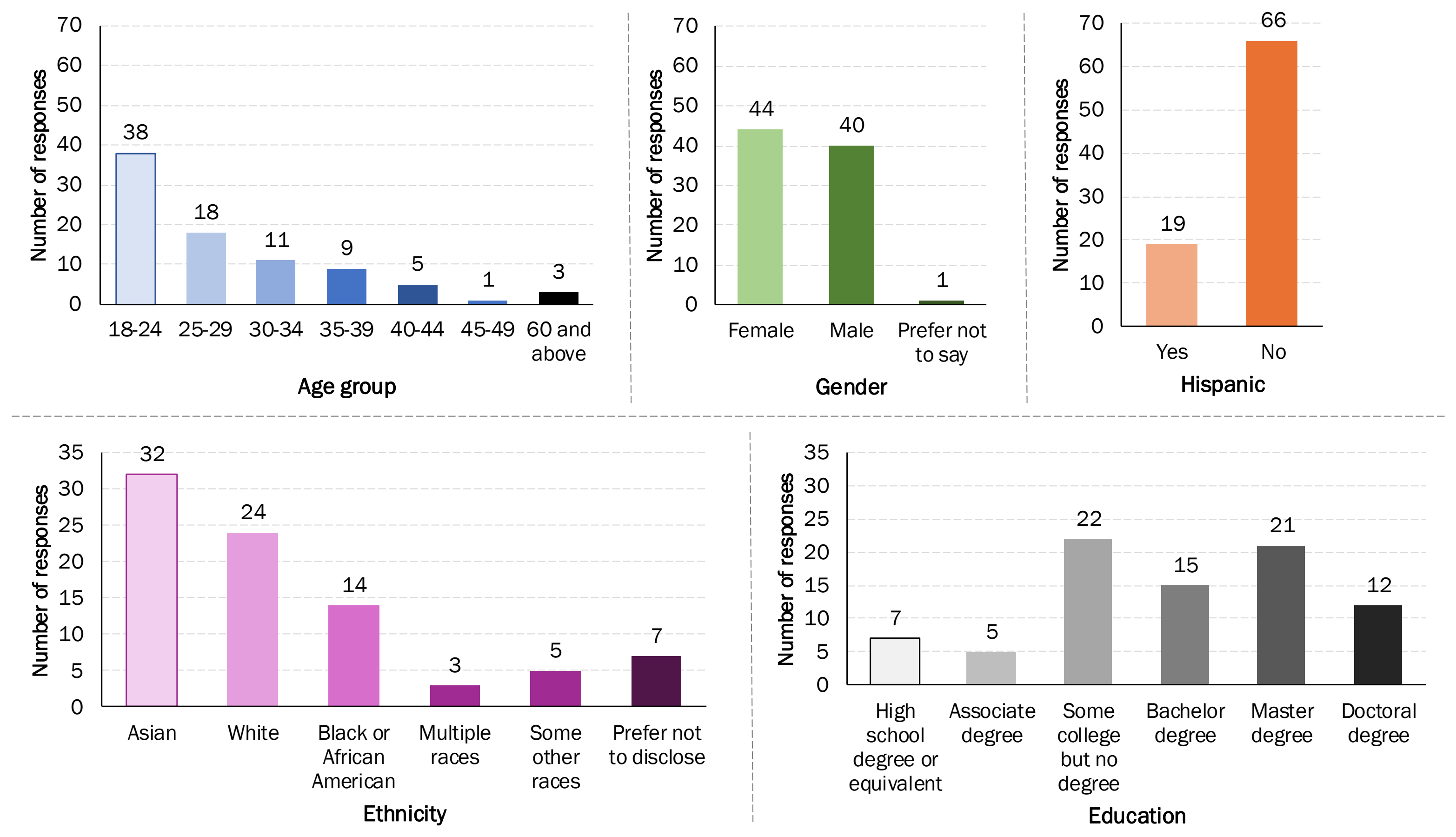}
\caption{Participant demographic information}\label{fig:A1}
\end{figure}

The study cohort was demographically well-balanced, with participants representing a broad mix of age groups, educational backgrounds, ethnicities, and gender, as shown in Figure~\ref{fig:A1}. This diversity enhanced the generalizability of the study and supported a comprehensive understanding of how users with varied backgrounds interact with LLM-integrated BEMS. We did not include this information since they were not included in the analysis.

\section{SCALE Framework: Prompt and Rubric}

\begin{figure}[tbp]
\centering
\includegraphics[width=0.80\textwidth]{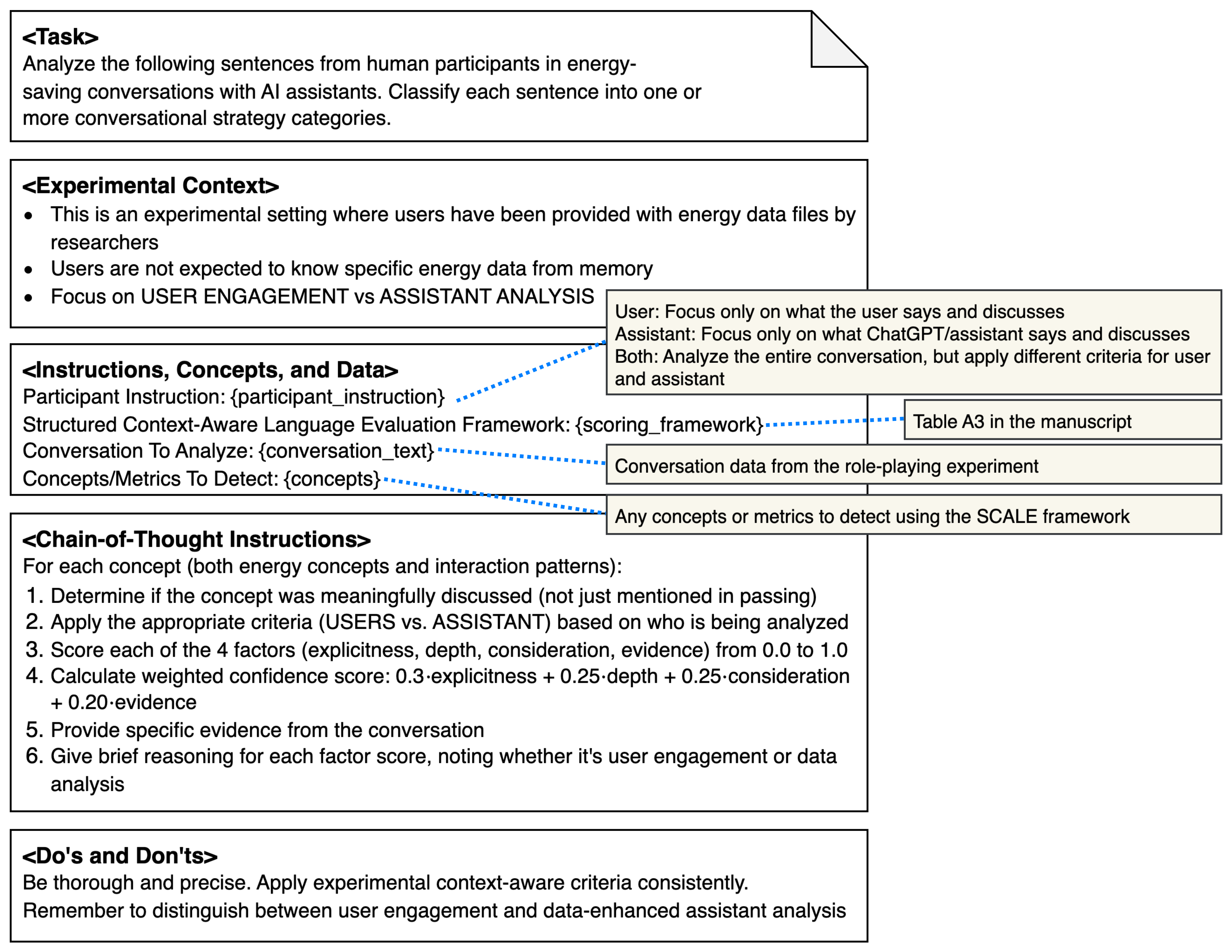}
\caption{Prompt for the SCALE framework}\label{fig:A2}
\end{figure}

We developed a prompt, as shown in Figure~\ref{fig:A2}, to implement the SCALE framework. The entire code for this GPT-based SCALE framework is shared in the following link:

\begin{itemize}
\item
  \url{https://github.com/humanbuildingsynergy/HUBS_GPT-based_SCALE_Framework}.
\end{itemize}

For the reproducibility of the SCALE framework, we developed scoring rubrics as organized in Table A3.

\smallertable
\begin{table}[p]
\centering\footnotesize\renewcommand{\arraystretch}{0.95}
\caption{Four scoring factors in the proposed context-aware multi-factor scoring framework.}\label{tab:A2}
\begin{tabularx}{\textwidth}{@{}>{\raggedright\arraybackslash}p{3.1cm}>{\raggedright\arraybackslash}X@{}}
\toprule
\multicolumn{2}{@{}l}{\textbf{Factor \#1: Explicitness}} \\
Description & How directly and clearly the concept is mentioned using appropriate terminology \\
Scoring criteria for user prompts & 1.0: Domain terminology used correctly in context\newline
0.8: General energy terms used appropriately in context\newline
0.6: Related terms that imply concept understanding\newline
0.4: Indirect references requiring some inference\newline
0.2: Vague connection or unclear terminology\newline
0.0: No mention or connection \\
Scoring criteria for GPT responses & 1.0: Precise domain terminology with technical accuracy (kWh, TOU rates, HVAC)\newline
0.8: Appropriate technical terms with clear explanations\newline
0.6: General energy terminology used correctly\newline
0.4: Basic energy terms with some accuracy\newline
0.2: Imprecise or unclear terminology\newline
0.0: No meaningful terminology used \\
\midrule
\multicolumn{2}{@{}l}{\textbf{Factor \#2: Depth}} \\
Description & Quality and authenticity of engagement with the concept \\
Scoring criteria for user prompts & 1.0: Deep contextual engagement with detailed reasoning, constraints, or comprehensive understanding\newline
0.8: Substantial contextual engagement with clear reasoning or thoughtful analysis\newline
0.6: Moderate contextual engagement with some reasoning or understanding\newline
0.4: Basic contextual engagement with minimal reasoning or surface-level understanding\newline
0.2: Shallow engagement with little reasoning or very limited understanding\newline
0.0: No meaningful engagement demonstrated \\
Scoring criteria for GPT responses & 1.0: Comprehensive multi-faceted analysis examining multiple dimensions of the concept\newline
0.8: Thorough analysis exploring several aspects or implications of the concept\newline
0.6: Moderate analysis covering key aspects with reasonable detail\newline
0.4: Basic analysis touching on main points with limited development\newline
0.2: Superficial analysis with minimal exploration or development\newline
0.0: No meaningful analytical processing demonstrated \\
\midrule
\multicolumn{2}{@{}l}{\textbf{Factor \#3: Consideration}} \\
Description & Whether the concept was meaningfully present in the conversation \\
Scoring criteria for user prompts & 1.0: Concept clearly influences user's decisions, preferences, or thinking\newline
0.8: Concept is actively considered in relation to user's situation\newline
0.6: Concept is acknowledged and shows contextual relevance\newline
0.4: Concept is mentioned with minimal contextual connection\newline
0.2: Concept is barely acknowledged or referenced\newline
0.0: Concept is not considered in user's thinking \\
Scoring criteria for GPT responses & 1.0: Concept is fully integrated into analysis and recommendations\newline
0.8: Concept is clearly incorporated into response strategy\newline
0.6: Concept is meaningfully addressed in the analysis\newline
0.4: Concept is mentioned but not well integrated\newline
0.2: Concept is briefly touched upon\newline
0.0: Concept is not considered in the response \\
\midrule
\multicolumn{2}{@{}l}{\textbf{Factor \#4: Evidence}} \\
Description & Quality and authenticity of supporting evidence provided \\
Scoring criteria for user prompts & 1.0: Contextual examples, specific constraints, or experimental details\newline
0.8: Clear situational context or constraints with details\newline
0.6: General contextual examples or reasonable situational context\newline
0.4: Some contextual information or basic examples\newline
0.2: Minimal supporting contextual information\newline
0.0: No contextual evidence or examples provided \\
Scoring criteria for GPT responses & 1.0: Multiple high-quality sources: quantitative data + domain expertise + specific examples\newline
0.8: Strong primary source with additional supporting information\newline
0.6: Solid single source with reasonable supporting details\newline
0.4: Basic source material with minimal additional support\newline
0.2: Weak or limited source material with little support\newline
0.0: No credible supporting information provided \\
\bottomrule
\end{tabularx}
\end{table}

\section{Energy Saving Strategy Comparison}

\begin{figure}[tbp]
\centering
\includegraphics[width=0.64\textwidth]{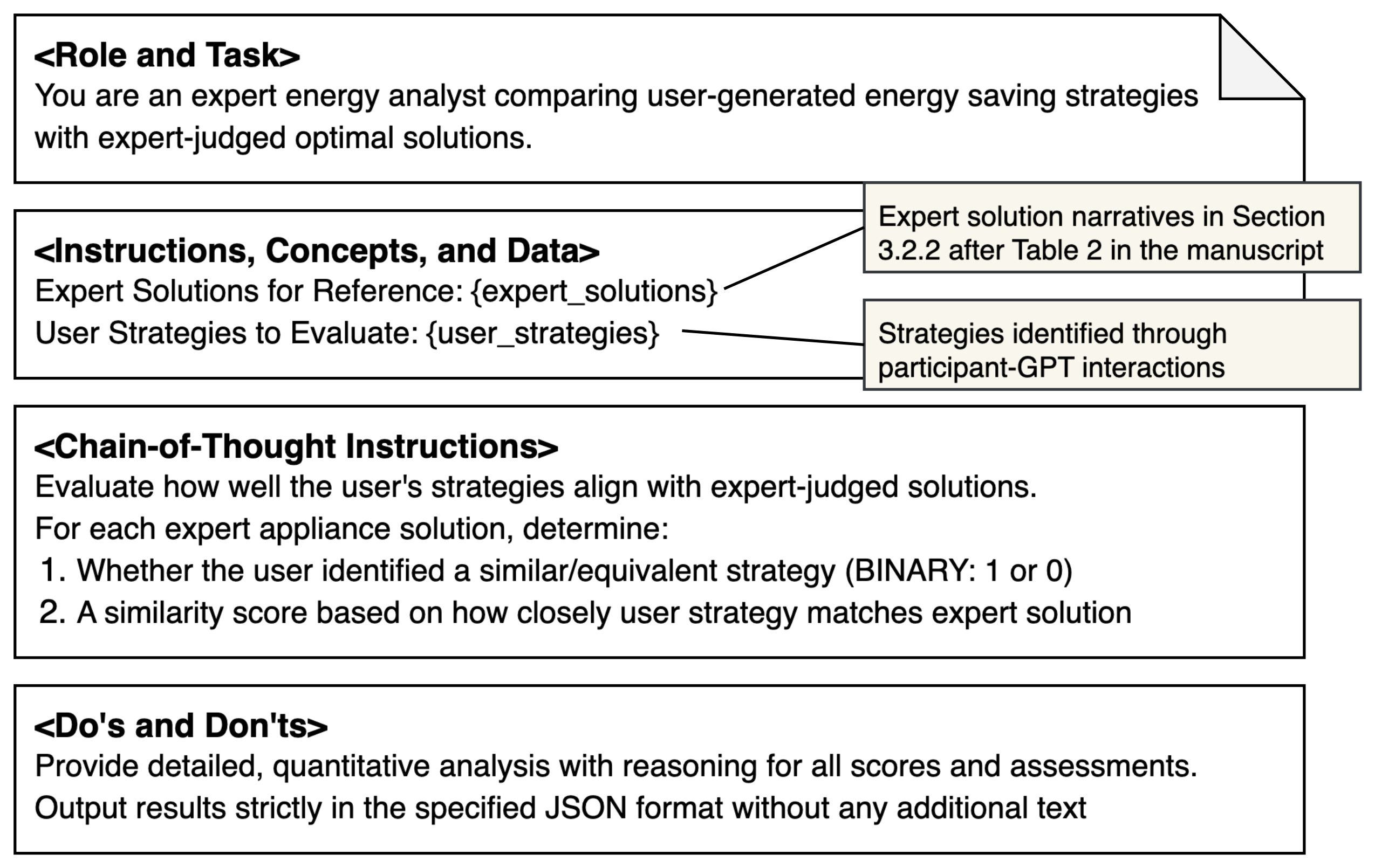}
\caption{Prompt to guide how two sets of strategies (user-GPT vs. reference) should be compared}\label{fig:A3}
\end{figure}

Figure~\ref{fig:A3} presents the prompt used to compare two sets of strategies drawn from human-AI interactions and expert judgments.

\section{Post-Experiment Survey}

\begin{surveyblock}
This post survey aims to understand any changes in your knowledge of home energy use, your expertise in GenAI tools like ChatGPT, and your perspectives on both the benefits of GenAI and its potential for managing home energy use.

Q1: Please rate your understanding of home energy use and its patterns after interactions with ChatGPT.

\begin{enumerate}
\def\labelenumi{\arabic{enumi})}
\item
  No understanding
\item
  Basic understanding
\item
  Moderate understanding
\item
  Good understanding
\item
  Advanced understanding
\end{enumerate}

Q2: Please rate your understanding of how monthly energy bills are calculated.

\begin{enumerate}
\def\labelenumi{\arabic{enumi})}
\item
  No understanding
\item
  Basic understanding
\item
  Moderate understanding
\item
  Good understanding
\item
  Advanced understanding
\end{enumerate}

Q3: How much do you agree with the following statement: I found it difficult to understand ChatGPT's answers because of my limited knowledge regarding home energy.

\begin{enumerate}
\def\labelenumi{\arabic{enumi})}
\item
  Strongly disagree
\item
  Disagree
\item
  Neither agree nor disagree
\item
  Agree
\item
  Strongly agree
\end{enumerate}

Q4: How much do you agree with the following statement: My interactions with ChatGPT have improved my knowledge of home energy use, its pattern, and the potential savings achievable through behavioral changes.

\begin{enumerate}
\def\labelenumi{\arabic{enumi})}
\item
  Strongly disagree
\item
  Disagree
\item
  Neither agree nor disagree
\item
  Agree
\item
  Strongly agree
\end{enumerate}

Q5: Please rate your level of expertise in using GenAI tools like ChatGPT after your role-playing experiment

\begin{enumerate}
\def\labelenumi{\arabic{enumi})}
\item
  Beginner
\item
  Lower intermediate
\item
  Intermediate
\item
  Upper intermediate
\item
  Expert
\end{enumerate}

Q6: How much do you agree with the following statement: I fully utilized ChatGPT's capabilities.

\begin{enumerate}
\def\labelenumi{\arabic{enumi})}
\item
  Strongly disagree
\item
  Disagree
\item
  Neither agree nor disagree
\item
  Agree
\item
  Strongly agree
\end{enumerate}

Q7: I wish I knew some techniques to enhance ChatGPT's responses.

\begin{enumerate}
\def\labelenumi{\arabic{enumi})}
\item
  Strongly disagree
\item
  Disagree
\item
  Neither agree nor disagree
\item
  Agree
\item
  Strongly agree
\end{enumerate}

Q8: How much do you agree with the following statement: Through my interactions with ChatGPT, I was able to arrive at better answers in the role-playing experiment

\begin{enumerate}
\def\labelenumi{\arabic{enumi})}
\item
  Strongly disagree
\item
  Disagree
\item
  Neither agree nor disagree
\item
  Agree
\item
  Strongly agree
\end{enumerate}

Q9: How much do you agree with the following statement: With ChatGPT, I could incorporate expert-level knowledge/analysis into my answers.

\begin{enumerate}
\def\labelenumi{\arabic{enumi})}
\item
  Strongly disagree
\item
  Disagree
\item
  Neither agree nor disagree
\item
  Agree
\item
  Strongly agree
\end{enumerate}

Q10: How much do you agree with the following statement: If I could manage my home energy usage, patterns and bills through interactions with ChatGPT -- similar to the role-play experiment -- I would do so.

\begin{enumerate}
\def\labelenumi{\arabic{enumi})}
\item
  Strongly disagree
\item
  Disagree
\item
  Neither agree nor disagree
\item
  Agree
\item
  Strongly agree
\end{enumerate}

Q11: How much do you agree with the following statement: If I could use GenAI tools like ChatGPT for my home energy use, I think my understanding of home energy use and its patterns will improve consistently and continuously.

\begin{enumerate}
\def\labelenumi{\arabic{enumi})}
\item
  Strongly disagree
\item
  Disagree
\item
  Neither agree nor disagree
\item
  Agree
\item
  Strongly agree
\end{enumerate}
\end{surveyblock}

\section{Representative Conversation Examples.}

Figures from A4 to A6 illustrate the overview of the conversations analyzed in Section 4.1.

\begin{figure}[tbp]
\centering
\includegraphics[width=1.00\textwidth]{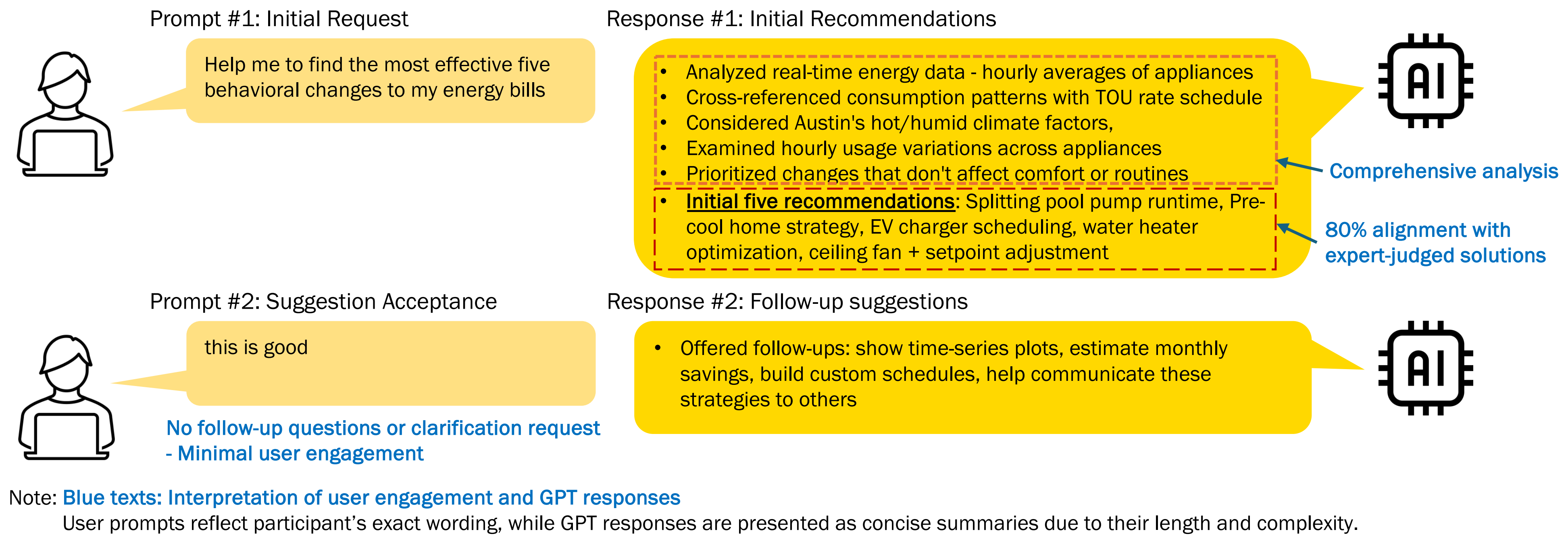}
\caption{Minimal user engagement, but high GPT performance in a participant-GPT collaboration}\label{fig:A4}
\end{figure}

\begin{figure}[tbp]
\centering
\includegraphics[width=1.00\textwidth]{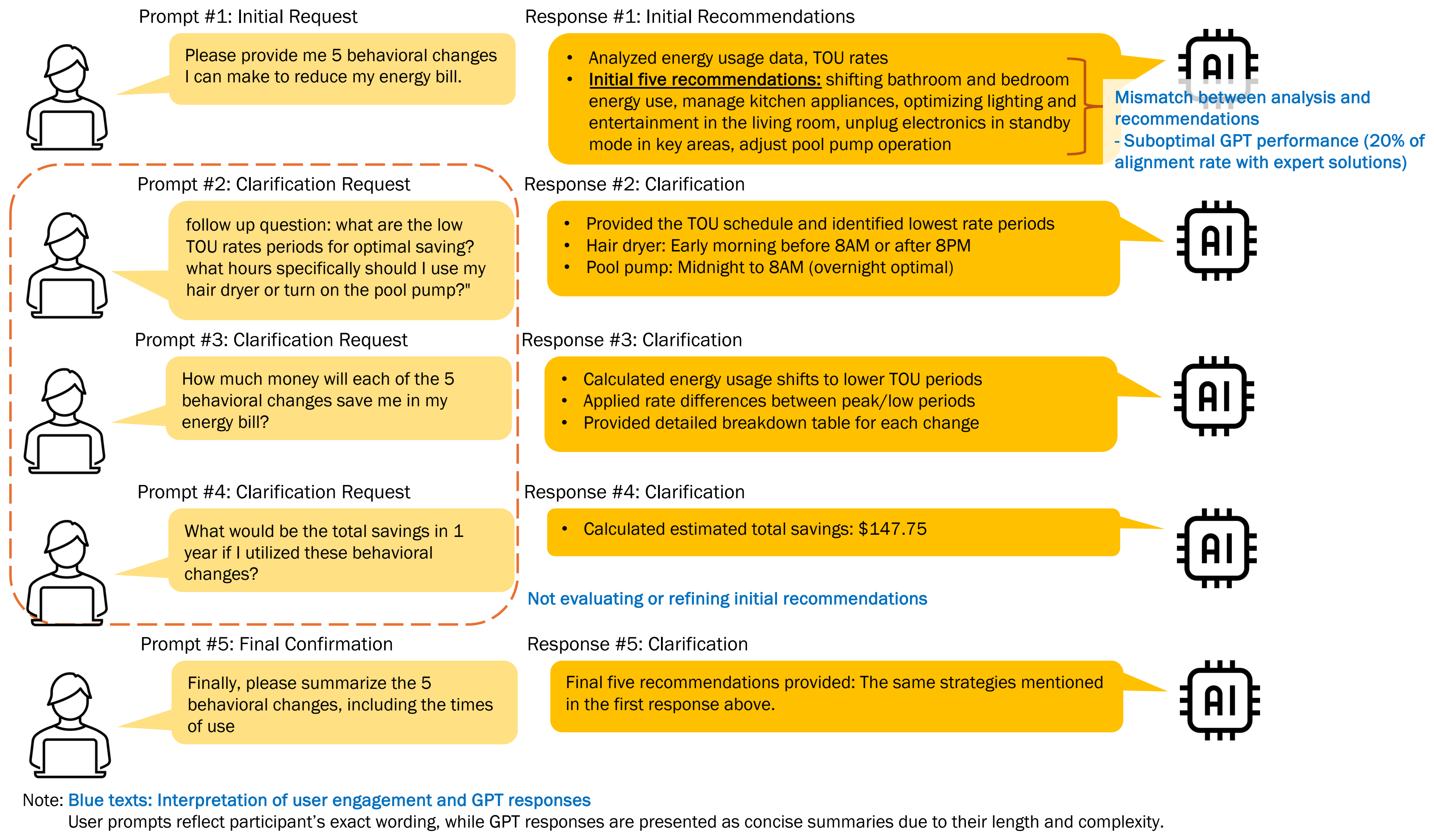}
\caption{Low to moderate user engagement and low GPT performance in a participant-GPT collaboration.}\label{fig:A5}
\end{figure}

\begin{figure}[tbp]
\centering
\includegraphics[width=1.00\textwidth]{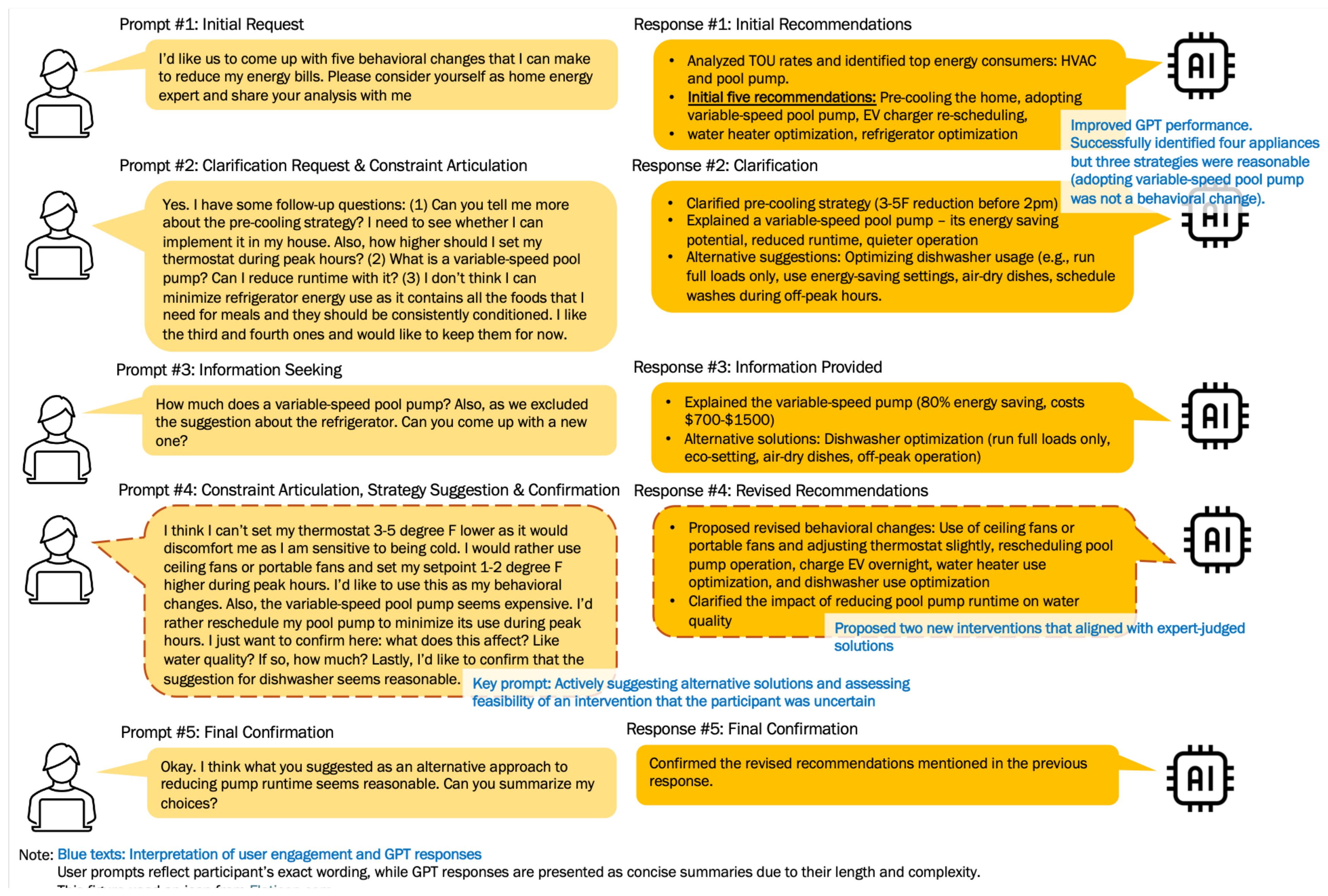}
\caption{High-quality, efficient user engagement and GPT's appropriate performance in a participant-GPT collaboration.}\label{fig:A6}
\end{figure}

\end{document}